\documentclass[lettersize,journal]{IEEEtran}
\IEEEoverridecommandlockouts
\usepackage{cite}
\usepackage{hyperref}
\usepackage{multirow}
\usepackage{amsmath,amssymb,amsfonts}
\usepackage{graphicx}
\usepackage{textcomp}
\usepackage{xcolor}
\usepackage{comment}
\usepackage{graphicx}
\usepackage{array}
\usepackage{xcolor}
\usepackage{tabularx}

\usepackage{makecell}
\usepackage{mathtools}
\usepackage{rotating}

\usepackage{amsmath,amsfonts}
\usepackage{array}
\usepackage[caption=false,font=normalsize,labelfont=sf,textfont=sf]{subfig}
\usepackage{textcomp}
\usepackage{stfloats}
\usepackage{url}
\usepackage{verbatim}
\usepackage{graphicx}
\usepackage{algorithmicx}
\usepackage{algorithm}
\usepackage{algpseudocode}

\def\BibTeX{{\rm B\kern-.05em{\sc i\kern-.025em b}\kern-.08em
    T\kern-.1667em\lower.7ex\hbox{E}\kern-.125emX}}
\begin{document}

%%%%%%%%%%%%%%%%%%%%%%%%%%%%%%%%%%%%%%%%%%%%%%%%%%%%%%%%%%%%%%%%%%%%%%%%%%%%%%%%
%%%%%%%%%%%%%%%%%%%              TITLE SECTION              %%%%%%%%%%%%%%%%%%%%
%%%%%%%%%%%%%%%%%%%%%%%%%%%%%%%%%%%%%%%%%%%%%%%%%%%%%%%%%%%%%%%%%%%%%%%%%%%%%%%%

%\title{Architectural Exploration of Application-Specific Resonant SRAM Compute-in-Memory (rCiM)}
\title{Time--to--Digital Converter (TDC)--Based Resonant Compute--in--Memory for INT8 CNNs with Layer--Optimized SRAM Mapping}

%\title{Low Power Cc-enhanced LC Resonant Clock Design Essentials}

%%%%%%%%%%%%%%%%%%%             AUTHORS SECTION             %%%%%%%%%%%%%%%%%%%%
\author{
%Anonymous for review purposes. Do not distribute.

Dhandeep Challagundla,~\IEEEmembership{Student Member,~IEEE},
%Ping~Y.~Lin,~\IEEEmembership{Student Member,~IEEE},
Ignatius~Bezzam,~\IEEEmembership{Member,~IEEE},
~and Riadul~Islam,~\IEEEmembership{Senior Member,~IEEE}
%<-this stops a space

\thanks{D. Challagundla and R Islam are with the Department 
of Computer Science and Electrical Engineering, University of Maryland, Baltimore County, 
MD 21250, USA e-mail: {riaduli@umbc.edu}.}
%\thanks{B Saha is with the Si2Chip Technologies, 
%Road 1B, Gayatri Tech Park, Bengaluru, Karnataka 560066, India e-mail: {biprangshu.saha@si2chip.com}}
\thanks{I Bezzam is with the Rezonent Inc., 
1525 McCarthy Blvd, Milpitas, CA 95035, USA e-mail: {i@rezonent.us}.}

\thanks{This research was funded in part by National Science Foundation (NSF) award number: 2138253, Rezonent Inc. award number: CORP0061, and UMBC Startup Fund.}
\thanks{Copyright (c) 2025 IEEE. Personal use of this material is permitted. 
However, permission to use this material for any other purposes must be 
obtained from the IEEE by sending an email to pubs-permissions@ieee.org.}
}
% The paper headers
%\markboth{IEEE Transactions on Circuits and Systems--I}
\markboth{IEEE JOURNAL ON EMERGING AND SELECTED TOPICS IN CIRCUITS AND SYSTEMS arXiv}
%\markboth{IEEE Transactions on Computer-Aided Design of Integrated Circuits and Systems}%
{Shell \MakeLowercase{\textit{et al.}}: ??????}

\newcommand{\fixme}[1]{{\Large FIXME:} {\bf #1}}

% make the title area
\maketitle

%%%%%%%%%%%%%%%%%%%             ABSTRACT SECTION            %%%%%%%%%%%%%%%%%%%%
\begin{abstract}
In recent years, Compute-in-memory (CiM) architectures have emerged as a promising solution for deep neural network (NN) accelerators. Multiply-accumulate~(MAC) is considered a {\textit de facto} unit operation in NNs. By leveraging the inherent parallel processing capabilities of CiM, NNs that require numerous MAC operations can be executed more efficiently. This is further facilitated by storing the weights in SRAM, reducing the need for extensive data movement and enhancing overall computational speed and efficiency. Traditional CiM architectures execute MAC operations in the analog domain, employing an Analog-to-Digital converter (ADC) to convert the analog MAC values into digital outputs. However, these ADCs introduce significant increase in area and power consumption, as well as introduce non-linearities. This work proposes a resonant time-domain compute-in-memory (TDC-CiM) architecture that eliminates the need for an ADC by using a time-to-digital converter (TDC) to digitize analog MAC results with lower power and area cost. A dedicated 8T SRAM cell enables reliable bitwise MAC operations, while the readout uses a 4-bit TDC with pulse-shrinking delay elements, achieving 1 GS/s sampling with a power consumption of only 1.25 mW. In addition, a weight stationary data mapping strategy combined with an automated SRAM macro selection algorithm enables scalable and energy-efficient deployment across CNN workloads. Evaluation across six CNN models shows that the algorithm reduces inference energy consumption by up to $8\times$ when scaling SRAM size from 32~KB to 256~KB, while maintaining minimal accuracy loss after quantization. The feasibility of the proposed architecture is validated on an 8~KB SRAM memory array using TSMC 28~nm technology. The proposed TDC-CiM architecture demonstrates a throughput of 320~GOPS with an energy efficiency of 38.46~TOPS/W.

\end{abstract}
\section{Introduction}

% Discuss the primary issues of ADC and integrating mixed-signal design with digital blocks
% Ensuring consistent performance across different operating conditions is a significant challenge.
% Verifying mixed-signal designs in complex systems, is a significant challenge. Testing analog circuits alongside digital components requires sophisticated testing methodologies.
% We can always talk about process variation affecting analog vs digital components. I am viewing our TDC as a digital component

\begin{figure}[t]
\begin{center}

\includegraphics[width = 0.5\textwidth]{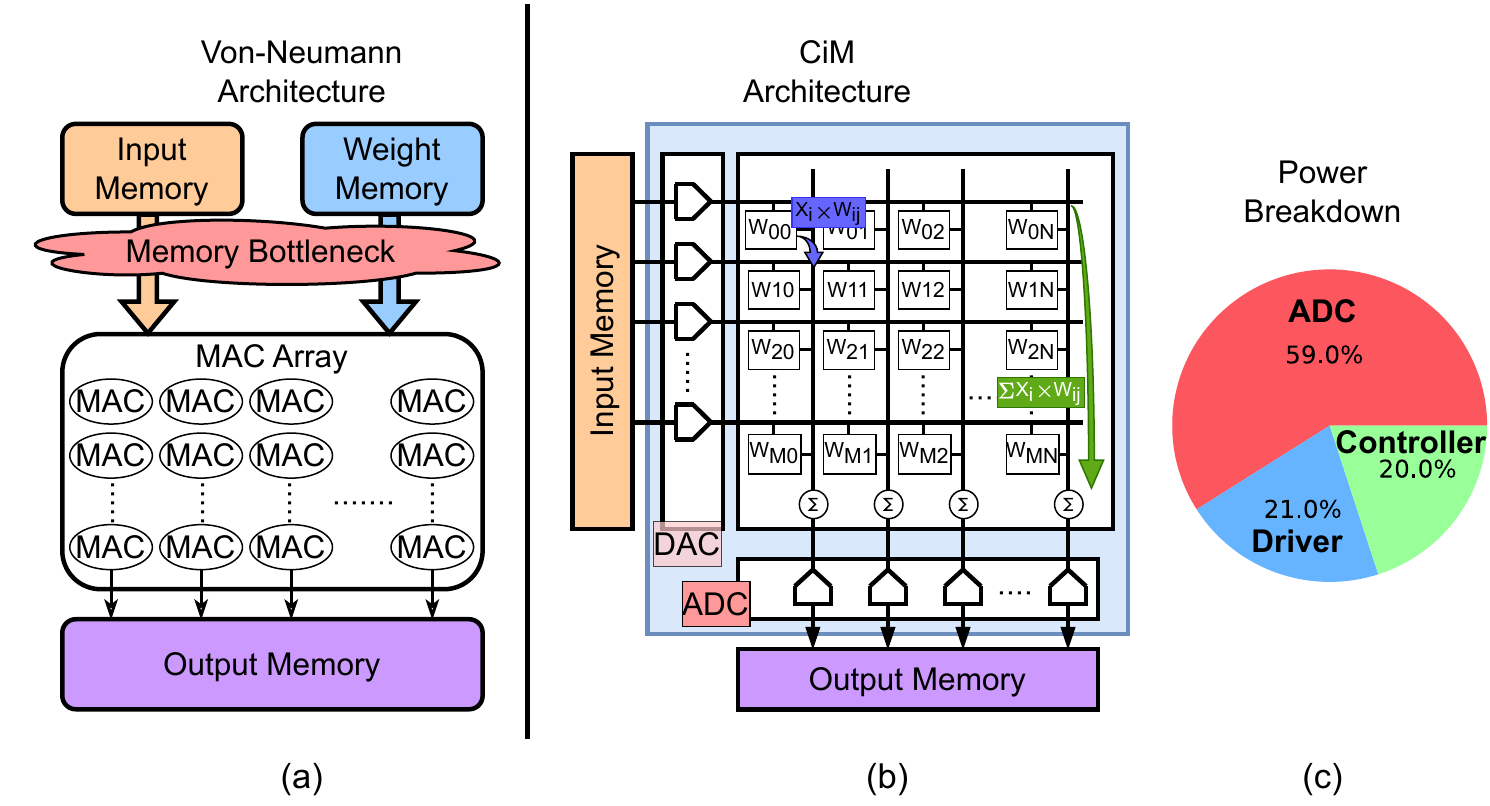}
\end{center}
%\vspace{-0.5cm}
\caption{(a) Traditional neural network accelerators use process elements for MAC computation, which causes memory wall bottlenecks due to high data movement. (b) CiM architectures mitigate the need for frequent data movements by enabling analog MAC computation within memory elements using charge accumulation and ADC converters, and (c) power breakdown of conventional CiM architectures shows 59\% of the total power consumption is caused by ADCs~\cite{ncim_Kim:23}.}

\label{fig:motivation}
%%\vspace{-0.50cm}
\end{figure}

Based on the Von Neumann architecture, neural network accelerators are currently being implemented in edge devices for complex tasks. These networks require substantial memory for accessing inputs and weights, creating memory wall bottlenecks that diminish the processor's performance, as shown in Figure~\ref{fig:motivation}(a). Computing-in-Memory (CiM) architectures reduce the energy overhead associated with data movement by leveraging parallel computation within DRAM~\cite{singh2022cidan,zhang_dram_23,singh_sram_25} and SRAM~\cite{choi_jetcas_22, li_sram_25, Lin_jssc:2021,  Wang:2019} memory arrays. Like conventional neural networks, the multiply-accumulate (MAC) operation is a fundamental process in CiM computations. CiM architectures execute several multiplications between inputs and weights concurrently, summing the results in the current domain. The voltage of the bitline corresponds to the final output of these MAC processes, as shown in Figure~\ref{fig:motivation}(b). Subsequently, an analog-to-digital converter (ADC) is utilized to convert the analog voltage into digital output bits. However, this analog processing, as shown in Figure~\ref{fig:motivation}(c), accounts for a prohibitively large amount (i.e., up to 59\%) of the total power consumption in a CIM architecture~\cite{ncim_Kim:23}.

Minimizing the overhead associated with ADCs is pivotal to improving the energy efficiency of CiM accelerators, unlike other energy-saving techniques~\cite{Gonugondla:2018, Islam_diff:2015, jyothi2025gate, Islam_neg:2021, Sun_iot:2025, guthaus2017current, Ji_sub:2025, Islam_cjece:2019, wan2025subthreshold, Islam_cm:2014}. Some CiM architectures achieve this by reducing the ADC precision for sparse inputs, whereas other approaches employ reduced-precision ADCs with non-linear quantization techniques~\cite{Saxena:22}. A primary issue in the integration of ADCs within mixed-signal design architectures, such as CiMs, is ensuring consistent ADC performance across diverse operating conditions and scaling with technology. Additionally, the verification of analog circuits along with digital components necessitates high-cost advanced testing methodologies due to the impact of process variations. 

To overcome the non-linearity and high power consumption of ADCs, more research is focused on Time-domain ADCs (TD ADCs) in which the analog voltage is converted to delay that can be processed in digital domain~\cite{tdc_comp_1,tdc_comp_4}. In this work, we alleviate the ADC issues in CiM computation by introducing an ADC-less Resonant time-domain CiM~(TDC-CiM) architecture using a time-to-digital converter (TDC) that performs the same functionality as an ADC while mimicking the behavior of a digital circuit to perform MAC operations within the SRAM memory elements~\cite{Challagundla_glsvlsi:2024}.
In particular, the main contributions of this work are:
\begin{itemize}
    \item First-ever ADC-less resonant CiM architecture for MAC operations utilizing a new TDC.
    \item A dedicated read-port 8T SRAM cell that enables read-disturb-free bitwise multiplications.
    \item An automated SRAM macro selection algorithm to map any given quantized CNN workload to the most energy-efficient TDC-CiM configuration by balancing kernel dimensions, required parallelism, and weight stationary mapping constraints.
    \item Functionality validation of the TDC and robustness analysis of TDC, demonstrating an energy efficiency of 163 fJ per conversion step.
\end{itemize}
%%\vspace{-0.2cm}

\section{Background}

% various CIM architectures are referecend. this showcases the different operations begin performed using the CIM 

As an emerging paradigm, SRAM-based CiM architectures have shown promising potential in significantly enhancing processing speed and energy efficiency for a wide range of computing tasks, such as MAC~\cite{mac_comp_1, Angizi:21, mac_comp_2, mac_comp_3, mac_comp_4}, CAM~\cite{Feng:23}, and boolean logic~\cite{Sunrui:23, Challagundla_iccd:2023}. Besides, both series~\cite{challagundla2023design, islam2024system, Challagundla_ms:2022} and parallel resonance~\cite{Islam_low:2018,friedman_resonance, islam2011high, mizukawa2024influence, Lin:2015} exhibit tremendous potential in energy-efficient computing. While series resonance allows a wide operating frequency range, this research deploys a parallel resonance scheme for the fixed operating frequency used in the SRAM characterization.

%% comparing the previous CIM techniques using different sram cells and their disadvantages. compared 6t which has read-disturb, 7t which has reverse current flow on readbitlines. 9T1C sram cell which has high latency for charging the capacitor when 1x1=1 operation.

CiM architectures utilizing standard 6T SRAM cells~\cite{mac_comp_1, mac_comp_2} perform computations by propagating information across the bitlines. This computational approach requires multi-row activations to fetch several operands within a single access cycle. However, this leads to high voltage changes along the bitlines, causing serious read-disturb issues potentially leading to data corruption.~\cite{mac_comp_4} employs a 7T SRAM cell for MAC computations by leveraging the discharge of the read bitline to eliminate read-disturb issues. However, this method may lead to reverse current flow if the voltage of the read bitline has significantly dropped, while one row might not discharge, it could instead cause an increase in voltage on the bitline, inducing non-idealities in MAC computation and affecting the accuracy. \cite{mac_comp_3} utilizes a 9T1C architecture for MAC operations but suffers from high latency, primarily due to the time consumed during charging the capacitor during bitwise multiplication operations. \cite{Feng:23} uses an 11T SRAM cell to perform Boolean and CAM operations while eliminating the need for column-wise computation constraints. However, it suffers from high area consumption associated with the additional transistors used in the SRAM cell. This work uses an 8T SRAM bitcell that provides full read bitline swing that avoids the reverse current issue reported in \cite{mac_comp_4}, and reduces the area overhead compared to the 11T bitcell in \cite{Feng:23}.

%% how CiM architectures perform MAC computation. Mainly 2 techniques all analog, or all digital. Their disadvantages and advantages over one another and what architecture is used in this work.

{ 
CiM architectures generally employ MAC operation using two design approaches: analog-based~\cite{mac_comp_3} and digital-computing~\cite{Heng:24}. The first type of CiM architectures executes MAC computations using current~\cite{Charles:18}, voltage~\cite{Biswas:19}, or pulse width~\cite{Song:21} to represent information. These architectures benefit from low power consumption but suffer from various non-idealities. Conversely, the computation results in the digital CiM architectures are accurate when multi-bit operations are applied but suffer from high power consumption. Charge domain CiM is effective because accumulation happens on capacitors and charge adds linearly with low loss, where mulitple rows can be activated while maintaining accuracy~\cite{cheon_23,song_23}. Using metal oxide metal capacitors that are stable across process, voltage, and temperature (PVT), these arrays support higher parallelism per operation than current domain designs and consume lower energy~\cite{samba_23,picoram_25}. The ADC at readout has been the limiting element in many prior systems. Increasing the resolution of ADC to obtain higher precision operation lowers quantization error but also narrows the 1 LSB voltage step. Hence the same noise occupies a larger fraction of a code, and ADC errors become more frequent. The analog front end of the ADC also dominates power and is hard to guard across corners, which increases the energy consumption and introduces partial sum errors that reduce the overall accuracy~\cite{comp_jssc_24}. To address these issues, this work introduces an ADC-less CiM architecture that performs the MAC computation and converts the analog MAC value into a digital output through digital TDC circuitry. The TDC-CiM uses low-energy computation using capacitor-based computing while avoiding the ADC cost by mapping the MAC voltage swing to time and digitizing with a compact TDC built from digital blocks. The TDC is co-designed with the MAC discharge range and provides a monotonic 4-bit readout that can be linearized with a small lookup table if needed, preserves INT8 inference accuracy, and lowers converter power by removing the high-resolution ADC from the loop.}

Numerous studies on TDC circuits have been documented in the literature. The TDC architecture in \cite{tdc:1997} employs a delay element incorporated with a delay-locked loop and a counter to count the number of pulses resulting in a digital output. Another approach \cite{tdc:1995} utilizes two pulse-shrinking delay lines and a delay stabilization loop to convert a pulse input into digital output. \cite{tdc_comp_4} uses a 2-step TDC and a voltage-to-time converter to produce a digital output for a given input. Unlike the previous TDC works, this work uses a simple pulse-shrinking delay element alongside DFFs to capture the pulse count depending on the voltage input, further encoded as digital bits. %The TDC architecture is further discussed in Section~\ref{sec:tdc}.

\section{Proposed Architecture}
In a convolutional neural network (CNN), the multiply-accumulate (MAC) operation is a fundamental computation. This paper introduces a novel compute-in-memory (CiM) architecture that enhances the standard SRAM cache by enabling MAC operations within the memory structure. The design, detailed in Figure~\ref{fig:proposed_cim}, integrates a conventional SRAM array with a binary-weighted capacitor array to perform analog MAC operations using charge-sharing for 8-bit inputs and 8-bit weights. A Readout circuit formed by proposed Time-to-Digital (TDC) converters converts the analog MAC values to 2 sets of 4-bit digital values that are summed up to form an 8-bit MAC output.

\subsection{Proposed TDC-CiM architecture}
\begin{figure}[h]
%\vspace{-0.30cm}
\begin{center}
\includegraphics[width = 0.5\textwidth]{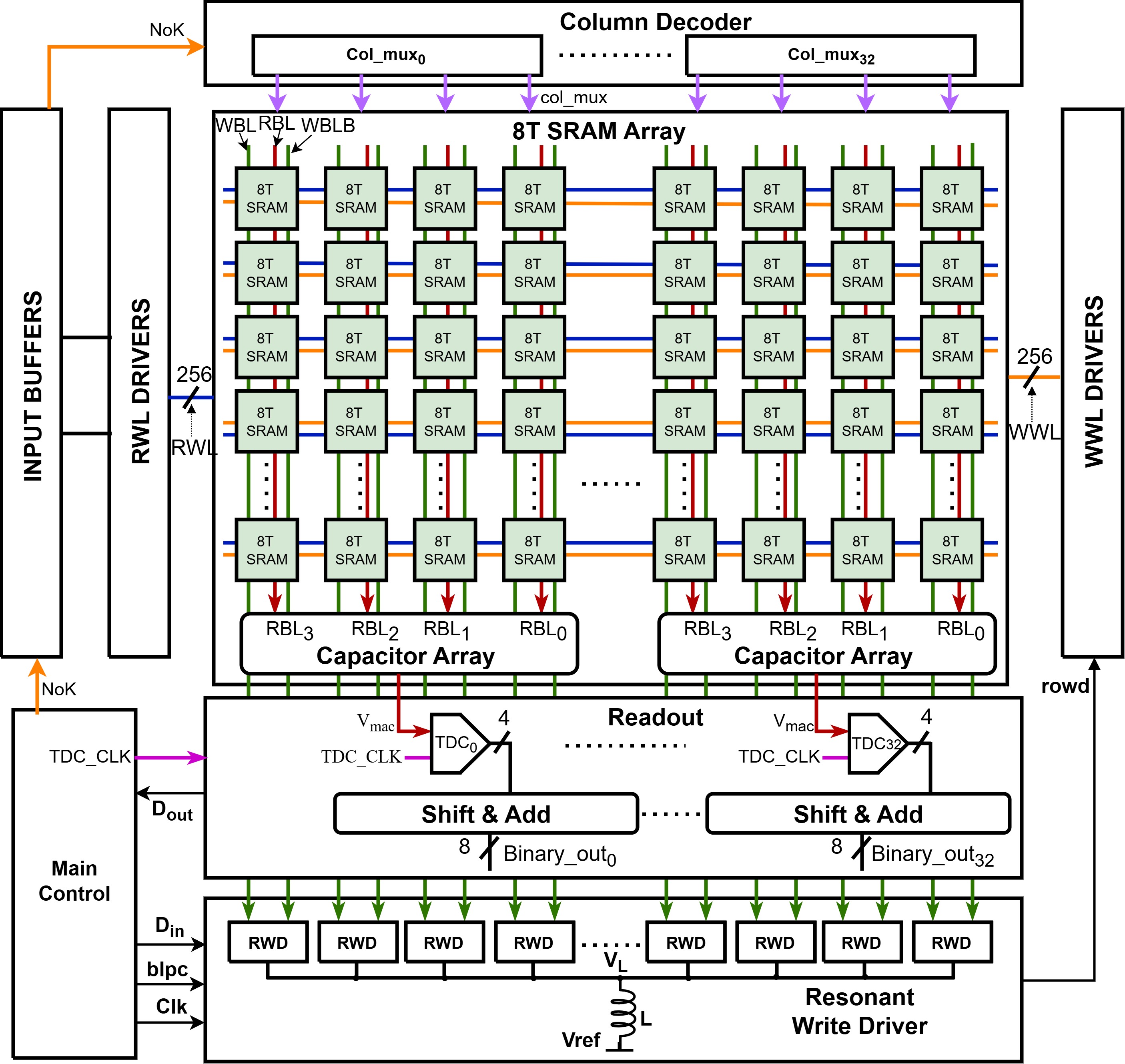}
\end{center}
%%%\vspace{-0.40cm}
\caption{The proposed architecture comprises the 8T bitcell array, RWL drivers to propagate the inputs, a new TDC-based readout circuit for executing MAC computations, and a resonant write driver for the writeback of MAC outputs.}
\label{fig:proposed_cim}
%%\vspace{-0.30cm}
\end{figure}

Figure~\ref{fig:proposed_cim} illustrates the architecture of the proposed TDC-CiM 8T SRAM macro, designed to facilitate MAC operations. This structure incorporates an 8T SRAM array, a binary-weighted capacitor array, and a readout circuit configured explicitly for MAC computations. The write-back of the MAC results is facilitated using the energy-recycling resonant write driver, enabling low-power write operations~\cite{rez_sram:21}. The architecture supports conventional read and write operations similar to traditional SRAM architectures with the help of row decoders for write wordlines~($WWL$) and read wordlines~($RWL$) and a peripheral control circuit to generate internal signals.

The 8T SRAM array consists of $256\times256$ bitcells to store and read using vertical bitlines and horizontal wordlines. During conventional read or write processes, the architecture is designed to activate either a single $RWL$ driver or a single $WWL$ driver at a time. However, nine $RWL$ drivers are simultaneously enabled to correspond with the $3\times3$ kernel of the input feature map~(IFM) for executing MAC operations. A weight stationary dataflow scheme is adopted in this work and further detailed in Section~\ref{sec:weight_stationary}. During MAC operations, the column decoder selects the appropriate columns containing the stored weights. The main control unit generates a signal indicating the number of kernels (NoK), which is then decoded to activate the corresponding columns. The 8-bit filter weights ($W$) are stored as logic ``1'' or ``0'' onto the bitcells, and the 8-bit $IFMs$ are applied in 2 cycles as a series of analog pulses on the $RWLs$.

\subsection{Proposed Multibit Multiplication}

\begin{figure}[h]
\begin{center}
\includegraphics[width = 0.5\textwidth]{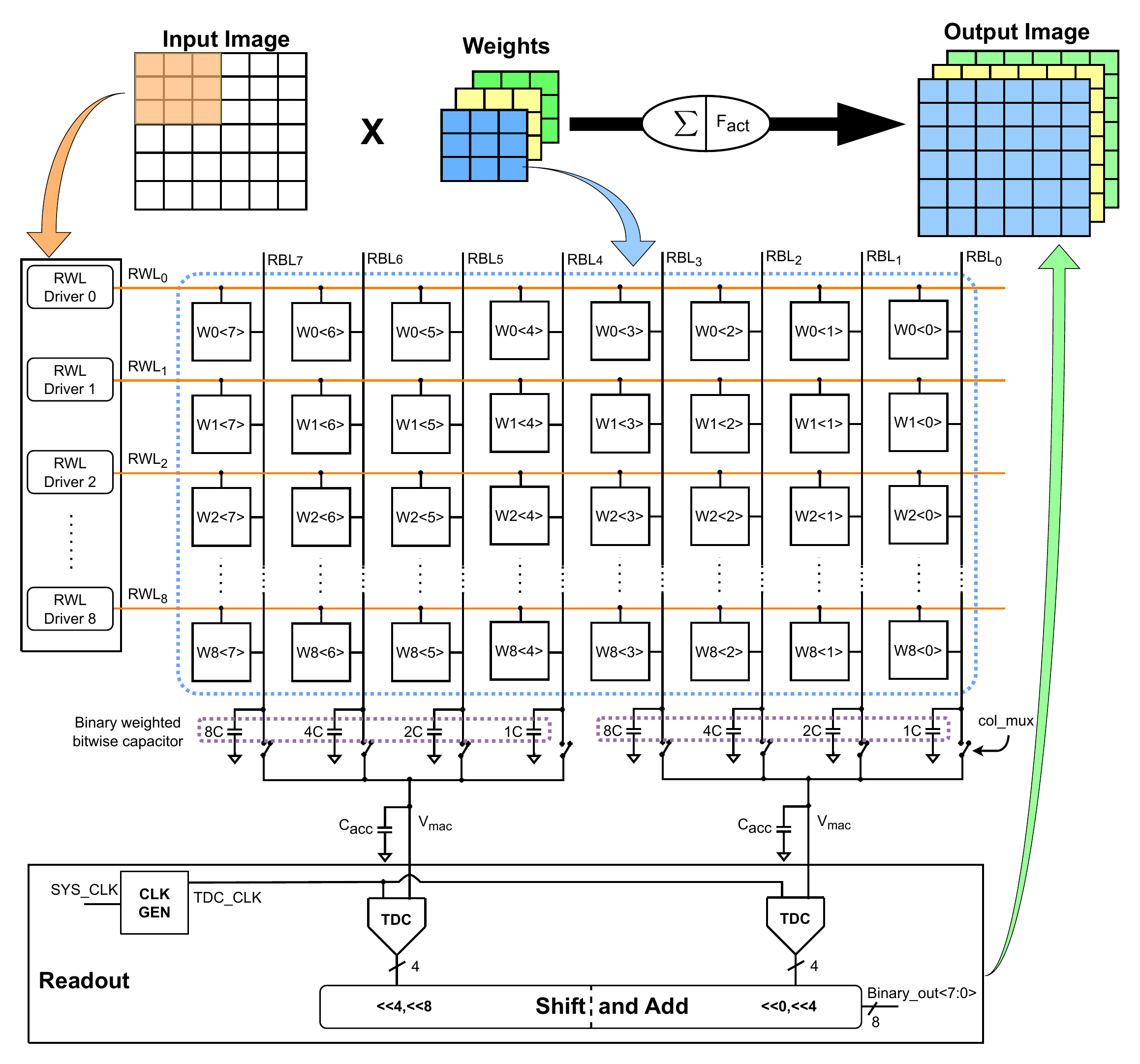}
\end{center}
%\vspace{-0.40cm}
\caption{Layer mapping of a convolution operation performed by applying input data through RWL drivers, multiplied with fixed weights in SRAM cells performing bitwise MAC operations along each bitline via binary-weighted capacitors, culminating in a multi-bit MAC output through charge sharing on the $C_{acc}$, where the resulting voltage $V_{mac}$ is digitized by the TDC.}
\label{fig:layer_mapping}
%\vspace{-0.30cm}
\end{figure}

In the proposed architecture, the 8-bit filter weights are loaded along the row direction in 2 sets of four adjacent 8T SRAM bitcells. The mapping of the filter weights inside the TDC-CiM array is shown in Figure~\ref{fig:layer_mapping}. The 8-bit $IFMs$ are applied in two subsequent clock cycles as a series of 4-bit input pulses into the TDC-CiM array for MAC operation.

\begin{figure}[h]
%%\vspace{-0.30cm}
\begin{center}
\includegraphics[width = 0.5\textwidth]{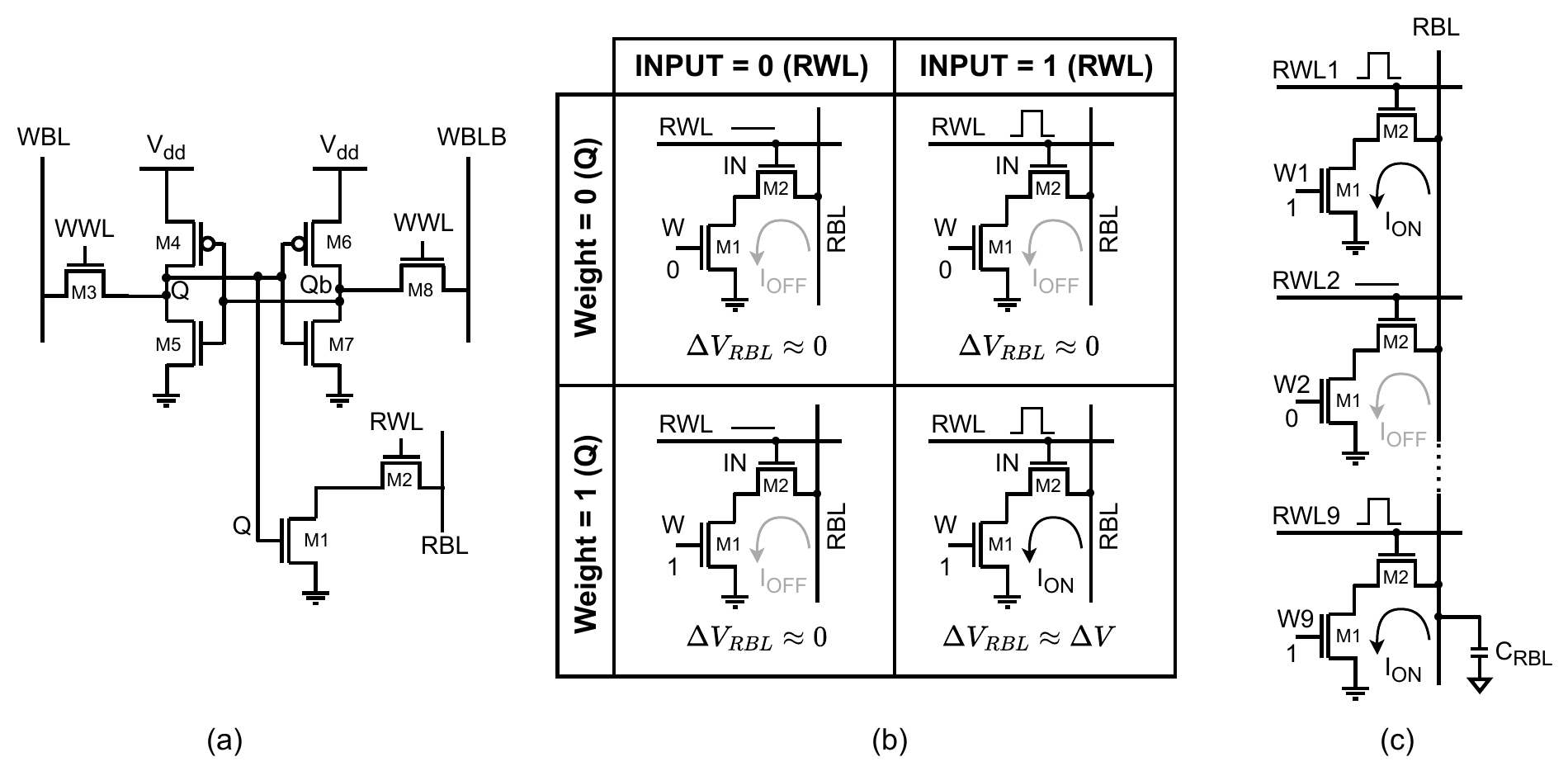}
\end{center}
%%\vspace{-0.40cm}
\caption{(a) Schematic of 8T SRAM bitcell with $M1-M2$ transistors forming a dedicated read port, (b) bitwise multiplication of inputs and weights using 8T bitcells show discharge of $\Delta V$ only when Input and Weight are ``1,'' (c) multirow activation executes series of bitwise multiplications that collectively perform a bitwise multiply-accumulate operation.}
\label{fig:bitwise_multiplication}
%%\vspace{-0.20cm}
\end{figure}

{\color{black}Figure~\ref{fig:bitwise_multiplication}(a) shows the transistor-level schematic of the 8T bitcell used for performing MAC operations. Figure~\ref{fig:bitwise_multiplication}(b) illustrates bitwise multiplication using an 8T SRAM bitcell. Here, INPUT = ``1'' is represented by a unit pulse, whereas INPUT = ``0'' indicates no pulse. When the weight (Q-value) is set to ``1,'' and the input is also ``1,'' a current flow occurs, discharging the RBL line by $\Delta V$. Hence, it performs a 1-bit multiplication of stored weight and input for a single bitcell. The voltage drop $\Delta V$ due to discharge can be captured by quantizing the amount of charge ($Q$) lost from the capacitor ($C_{RBL}$). The remaining charge on the capacitor can be written as:

\begin{equation}
Q_{\text{remaining}} = Q_{\text{initial}} - Q_{\text{discharged}}
\end{equation}

Since $Q = CV$, the remaining voltage ($V_{\text{rem}}$) on the capacitor ($C_{RBL}$) is:

\begin{equation}
V_{\text{rem}} \cdot C_{RBL} = (V_{\text{dd}} \cdot C_{RBL}) - (I_{\text{ds}} \cdot T_{\text{dis}})
\end{equation}

where, $I_{\text{ds}}$ is the current dischared for time period $T_{\text{dis}}$. Solving for $V_{\text{rem}}$:

\begin{equation}
V_{\text{rem}} = V_{\text{dd}} - \frac{I_{\text{ds}} \cdot T_{\text{dis}}}{C_{RBL}}
\end{equation}

Thus, the voltage drop $\Delta V$ can be quantized as:

\begin{equation}
\Delta V = V_{\text{dd}} - V_{\text{rem}} = \frac{I_{\text{ds}} \cdot T_{\text{dis}}}{C_{RBL}}
\end{equation}

%% \todo{By enabling several sounds uncertain}
Enabling multiple rows simultaneously lets us perform a series of these one-bit multiplications in parallel, leading to a collective discharge on the $RBL$ that represents the sum total of the individual multiplication operations, as shown in Figure~\ref{fig:bitwise_multiplication}(c). This process essentially accomplishes a bitwise MAC function across the RBL for the activated bitcells. Using 8T SRAM bitcells will overcome the limitation of reverse charging current in 6T/7T bitcells even when the bitwise multiplication yields a ``0." 

The total voltage drop on the $RBL$ due to $n$ parallel active rows can be expressed as:

\begin{equation}
\Delta V_{\text{total}} = \left( \sum_{i=0}^{n} X_i \cdot W_i \right) \cdot \left( \frac{I_{\text{ds}} \cdot T_{\text{dis}}}{C_{RBL}} \right)
\end{equation}

where $X_i~\&~W_i$ represent the input features and kernel weights, respectively. The bitwise MAC results are sampled on two sets of capacitors: binary-weighted bitwise capacitors and the accumulation capacitor $C_{acc}$, as shown in Figure~\ref{fig:layer_mapping}. The binary-weighted bitwise capacitors are discharged based on the number of rows activated. Once the bitwise multiplication is completed, the $col\_mux$ signal is turned ``ON,'' enabling charge-sharing between the bitwise capacitors and the $C_{acc}$. The $C_{acc}$ output node is the $V_{mac}$ voltage corresponding to the analog MAC result of the inputs and weights.}

%%%%% edit later

\begin{figure}[h]
\begin{center}
\includegraphics[width = 0.5\textwidth]{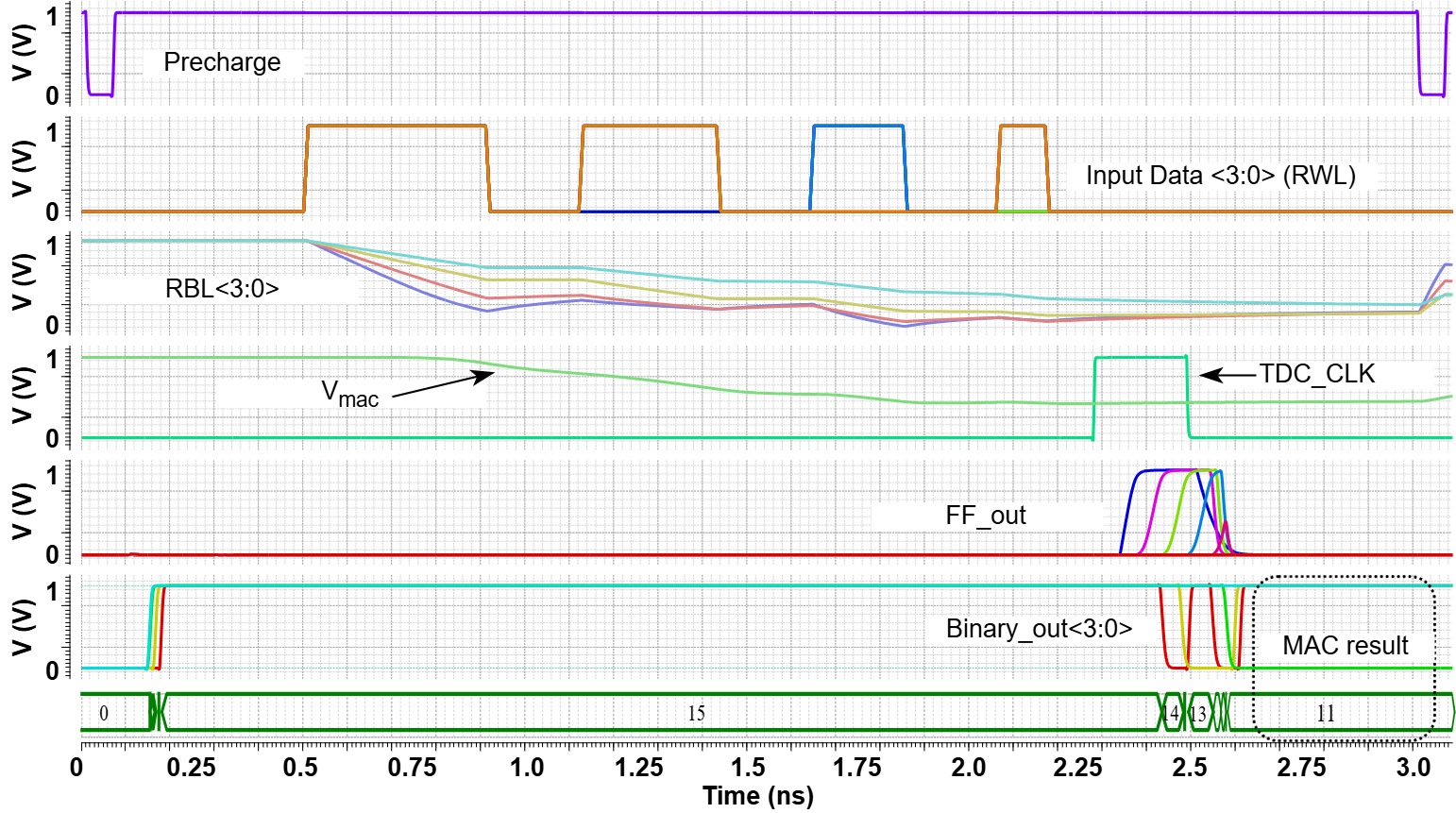}
\end{center}
%\caption{Functionality waveform illustrating NAND computation with "01"/"10" data and a successful energy-recycling writeback operation of the result.}
%\vspace{-0.50cm}
\caption{{The functionality simulation of performing a bitwise MAC operation on $RBLs$ and charge-sharing resulting in analog MAC voltage $V_{mac}$ which is digitized using the TDC.}}
\label{fig:mac_sim}
%\vspace{-0.30cm}
\end{figure}

Figure~\ref{fig:layer_mapping} illustrates the process of layer mapping in a CNN using an 8T SRAM cell array for performing MAC operations. The 8-bit $IFM$ is provided as input to the RWLs of the SRAM cells by the RWL drivers. A logical ``1'' in the input feature map generates a unit pulse on the RWL, while a logical ``0'' results in no pulse. Figure~\ref{fig:mac_sim} shows the SPICE simulation results for this MAC operation..

The kernel weights, which are 8-bit stationary values, are stored horizontally adjacent to 8T SRAM bitcells within the same row. Each RBL is initially connected to a binary-weighted capacitor. During the multiplication phase, the initial precharged RBLs discharge incrementally, depending on the number of activated rows. The configuration depicted consists of nine wordlines corresponding to a $3\times 3$ convolutional kernel typically used in CNN layers.
Once the bitwise multiplication process is completed, we turn ``ON'' the $col\_mux$ signal that connects the binary-weighted capacitors to the $C_{acc}$. This initiates a charge-sharing mechanism that charges the $C_{acc}$ capacitor based on the combined charge from the binary-weighted capacitors, producing the final accumulated voltage $V_{mac}$, as shown in Figure~\ref{fig:mac_sim}.

The accumulated voltage $V_{mac}$ in $C_{acc}$ represents the MAC operation's result in analog form. This analog value $V_{mac}$ is fed into a TDC along with the $TDC\_CLK$ pulse. The TDC translates this $V_{mac}$ voltage into a 4-bit digital output~($Binary\_out<3:0>$), representing the partial MAC value, which in this case is ``11.'' 

In the proposed scheme, the 4-bit input sequence is multiplied with two sets of 4-bit weights, producing two separate 4-bit partial MAC values. These partial MAC results are combined over two clock cycles. In the first clock cycle, the lower 4 bits of the IFM are processed, and the corresponding partial outputs are left-shifted by 0 and 4 bits, respectively. In the second clock cycle, the remaining upper 4 bits of the IFM are processed in the same manner, with the partial outputs left-shifted by 4 and 8 bits, respectively. The shifted partial sums from both cycles are then added together to generate the final 8-bit MAC result. This final sum is stored in an output buffer, which is then written back into the SRAM array using a resonant write driver~\cite{dhandeep_tvlsi} for use in the next layer.

\begin{figure}[h]
%\vspace{-0.50cm}
\begin{center}
\includegraphics[width = 0.42\textwidth]{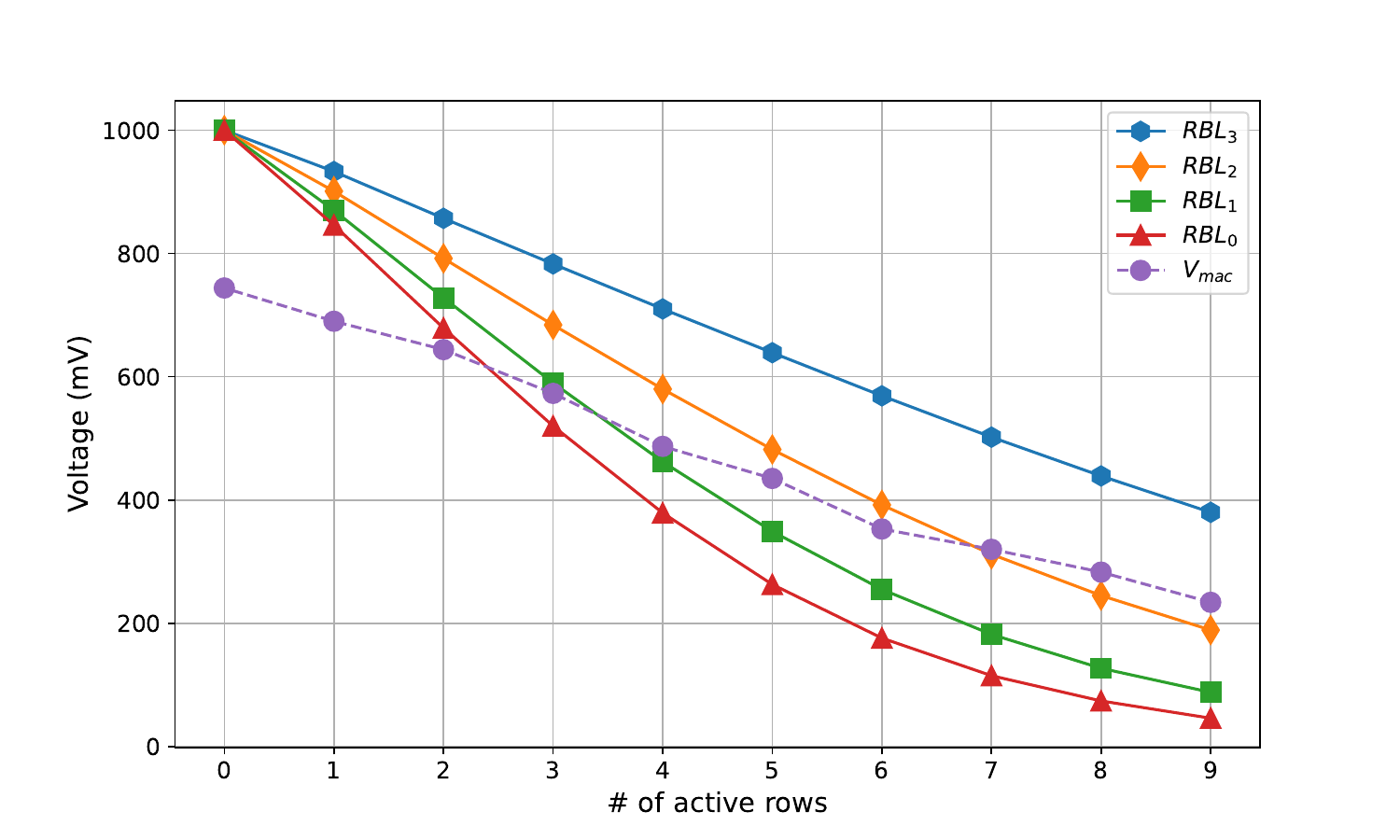}
\end{center}
%\caption{Functionality waveform illustrating NAND computation with "01"/"10" data and a successful energy-recycling writeback operation of the result.}
%\vspace{-0.50cm}
\caption{Discharge rates of the read bitlines (RBLs) and the accumulation capacitor $C_{acc}$ demonstrate a linear decrease in voltage with increasing number of rows activated.}
\label{fig:caps_discharge}
%\vspace{-0.40cm}
\end{figure}

The simulated discharge rate of the RBLs and the $C_{acc}$ relative to the number of active rows is shown in Figure~\ref{fig:caps_discharge}. As the number of rows activated increases, the discharge of both the RBLs and the accumulation capacitor also increases.

%%%%%%%%%%%%%%%%%%%%%%%%%%%%%%%%%%%%%%%%%%%%%%%%%%%%%%%%%%%%%%%%%%%%%%%%%%%%%%%%%%%%%%%%%%%%%%%%%%%%%%%%%%%%%%%%%%%%%%%%%%%%%%%%%%%%%%%%%%%%%%%%%%%%%%%%%%%%%%%%%%%%%%%%%%%%%%%%%%%%%%%%%%%%%%%
%%% end of MAC subsection

\subsection{Proposed TDC Architecture}
\label{sec:tdc}

\begin{figure}[h]
%%\vspace{-0.3cm}
\begin{center}
\includegraphics[width = 0.5\textwidth]{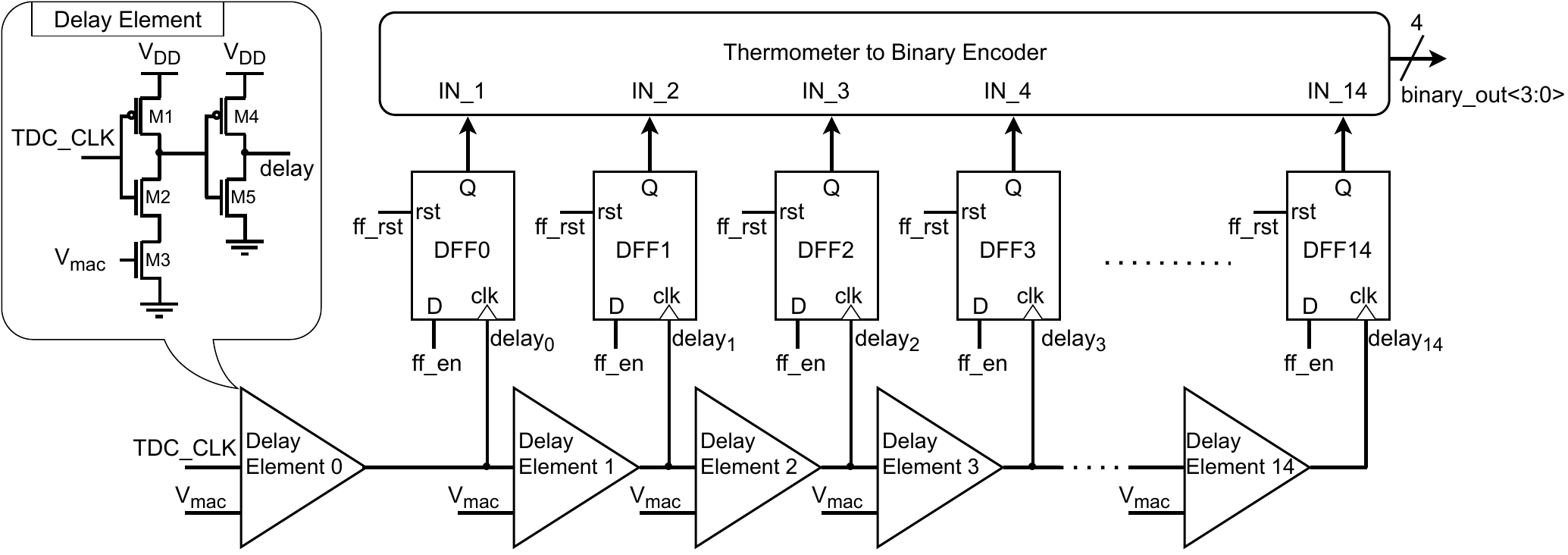}
\end{center}
%\caption{Functionality waveform illustrating NAND computation with "01"/"10" data and a successful energy-recycling writeback operation of the result.}
%\vspace{-0.30cm}
\caption{The 4-bit TDC converter architecture comprises an array of pulse-shrinking voltage-controlled delay elements along with DFFs to generate a pulse count corresponding to the analog MAC value $V_{mac}$, converted into digital output using a MUX-based thermometer-to-binary encoder.}
\label{fig:tdc_architecture}
%\vspace{-0.30cm}
\end{figure}

\begin{figure}[h]
%\vspace{-0.4cm}
\begin{center}
\includegraphics[width = 0.5\textwidth]{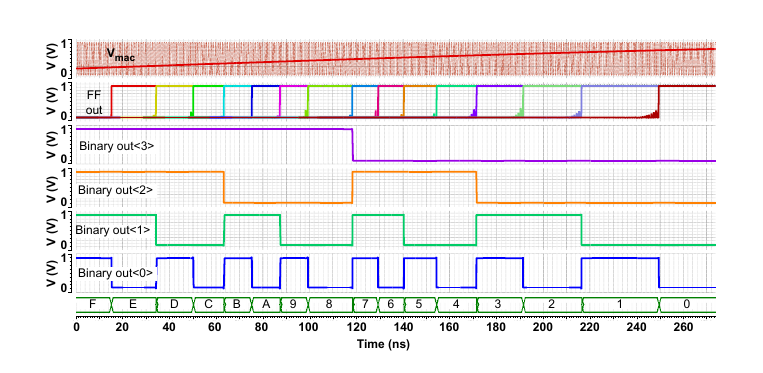}
\end{center}
%\caption{Functionality waveform illustrating NAND computation with "01"/"10" data and a successful energy-recycling writeback operation of the result.}
%\vspace{-0.50cm}
\caption{{The functionality simulation of the TDC shows a linear decrease of the digital output corresponding to the input voltage ramp signal $V_{mac}$, successfully capturing the full spectrum of 4-bit values.}}
\label{fig:tdc_functionality}
%%\vspace{-0.30cm}
\end{figure}

Figure~\ref{fig:tdc_architecture} illustrates the proposed 4-bit TDC. It comprises an array of pulse-shrinking delay elements alongside their corresponding D flip-flops ($DFFs$). The output ($Q$) of these $DFFs$ serve as inputs to the thermometer-to-binary encoder, resulting in the 4-bit binary output. The overall spice simulation of the TDC circuit is shown in Figure~\ref{fig:tdc_functionality}. The TDC takes the analog output from the MAC operation, which is stored as a voltage ($V_{mac}$) in the $C_{acc}$, and turns it into a digital signal. The TDC also has an input pulse, $TDC\_CLK$, generated from the system clock using a buffer-based delay circuit.

The transistors $M1-M5$, shown in Figure~\ref{fig:tdc_architecture}, form the pulse-shrinking delay element, which consists of the voltage-controlled buffer. The propagation of the rising edge of the
input pulse~($TDC\_CLK$) is slowed down by the current-starving transistor $M3$ while the falling edge travels fast. The $V_{mac}$ result obtained from the charge sharing after bitwise multiplication is provided as the voltage control input to the TDC circuit. Each delay element will shrink the input pulse by $\Delta T$ depending upon the $V_{mac}$ result. As the pulse travels through the delay line, the width of the pulse shrinks in each element by $\Delta T$ until the pulse entirely disappears.

Each pulse-shrinking delay element is connected to the clock pin of a positive edge-triggered $DFF$. The $DFF$ latch a ``1'' using the propagated pulse until the pulse disappears, which results in the following $DFF$ to retain a ``0.'' The simulation depicted in Figure~\ref{fig:tdc_functionality} captures the voltage required to set all the $DFFs$ to ``1'' by applying a voltage ramp signal.

The $Q$ from the $DFFs$ is subsequently provided as input to a MUX-based thermometer-to-binary encoder. This converter takes the string of ``1s'' and ``0s'' and converts them into a 4-bit binary number. This binary number is the digital version of the original analog voltage. This 4-bit digital output is provided as input to the shift\&add module for the final 8-bit MAC output computations.

{

\subsection{Resonant Write Driver}

The write path, shown in Figure~\ref{fig:res_driver}, uses an energy-recycling resonant driver with supply boosting following prior work~\cite{dhandeep_tvlsi,Islam_sram:2021}. A series inductor is placed in the discharge path of write bitlines \(Wbl/Wblb\) and with the other end connected to a reference voltage $Vref$. Choosing \(V_{\mathrm{ref}}=V_{\mathrm{DD}}/2\) maximizes the energy savings from the series inductor~\cite{dhandeep_tvlsi}. During a write operation, either \(Wbl\) or \(Wblb\) switches from logic “1” to logic “0,” the charge that would otherwise be dissipated is transferred and stored at \(V_{\mathrm{ref}}\) node. During the subsequent precharge phase, this stored energy is returned from \(V_{\mathrm{ref}}\) to the bitlines, which reduces the energy drawn from the supply and yields zero net current over the write–precharge cycle.

The resonant write driver in Figure~\ref{fig:res_driver} uses transistors $M5-M8$ to enable resonance by conditionally connecting the $Wbl/Wblb$ to the inductor controlled by the $vsr$ signal derived from the system clock. During the write of logic ``1" case shown in Figure~\ref{fig:res_driver_sim}, the $vsr$ signal enables the discharge path of the $Wblb$ signal for writing a data of ``1." The $vdn$ signal enables complete discharge of the bitline. During the subsequent precharge phase, \(vsr\) is asserted again so the energy stored in the inductor is returned to the bitlines. As a result, the precharge enable ($BLPC$) does not need to charge the bitline from 0V, reducing the overall energy consumption.

\begin{figure}[h]
%\vspace{-0.4cm}
\begin{center}
\includegraphics[width = 0.5\textwidth]{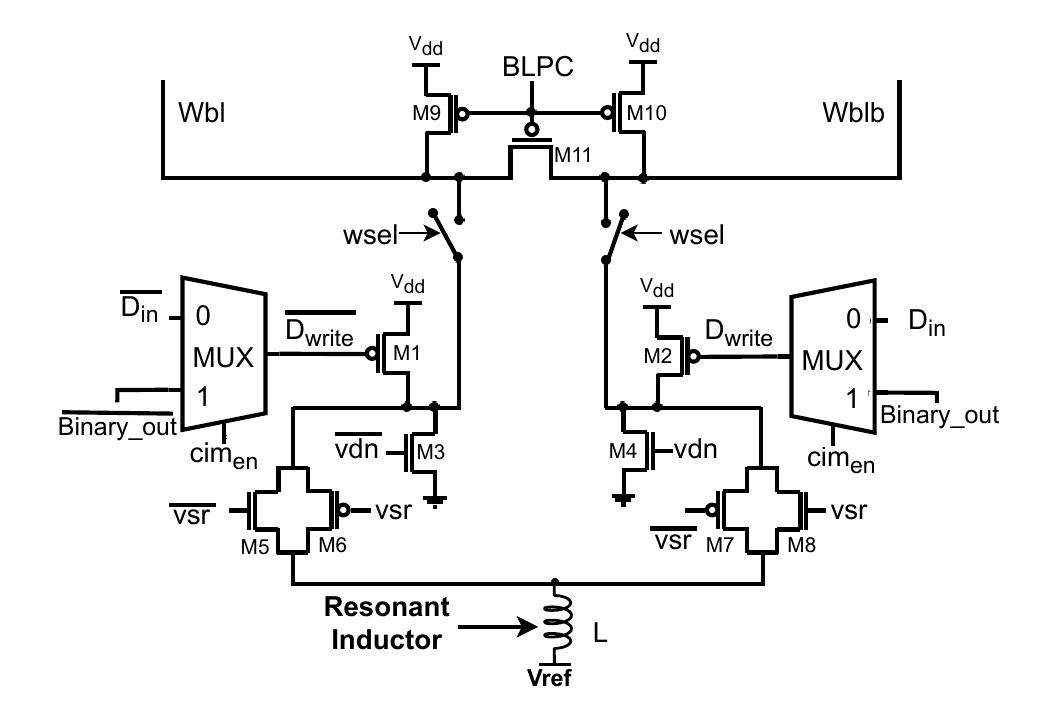}
\end{center}
\caption{{The resonant write driver has 4 additional transistors (M5 to M8) and a shared inductor over the traditional write driver to enable energy recycling during every write operation.}}
\label{fig:res_driver}
\end{figure}

\begin{figure}[h]
%\vspace{-0.4cm}
\begin{center}
\includegraphics[width = 0.35\textwidth]{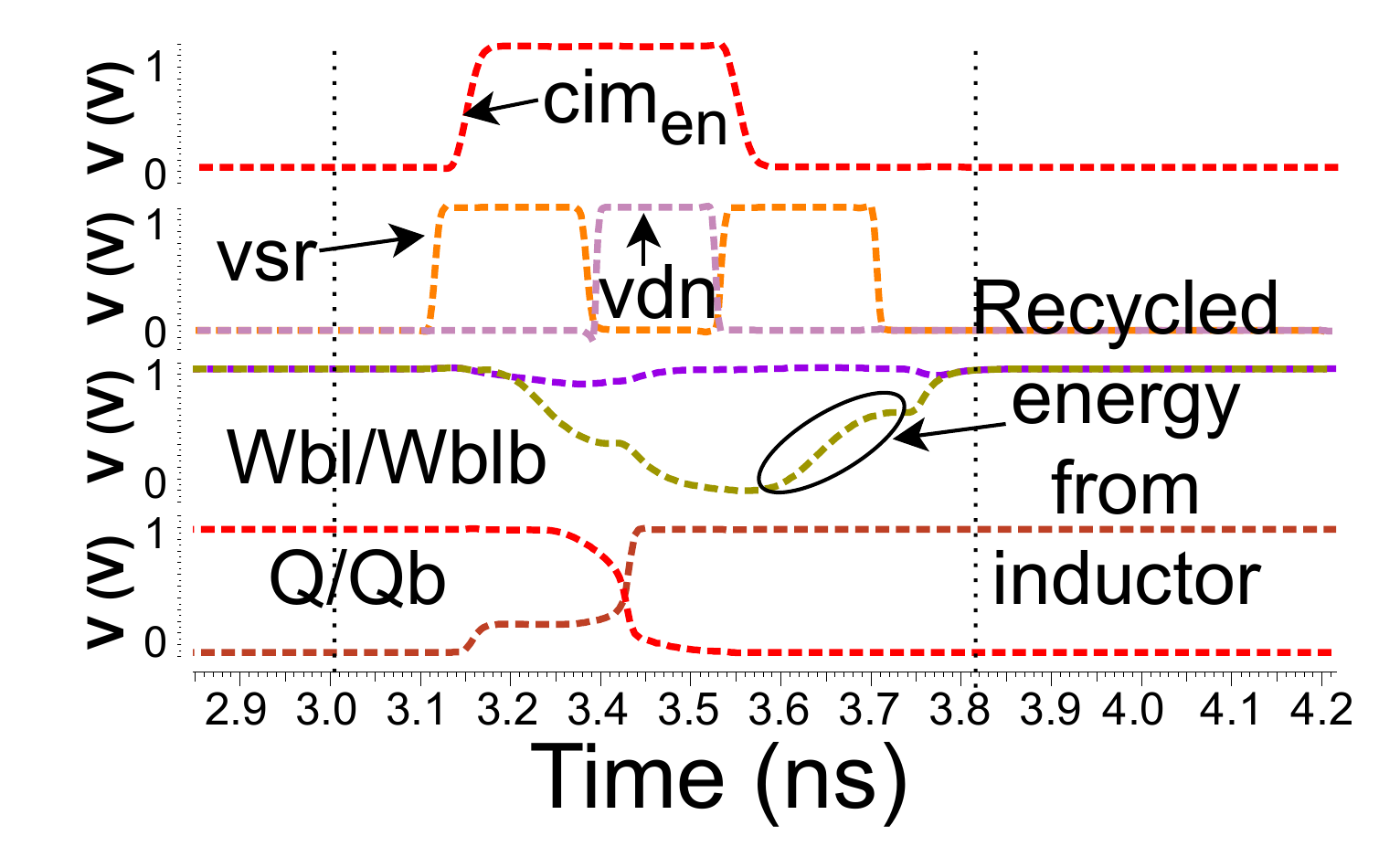}
\end{center}
\caption{{Functionality validation of the resonant write driver showcasing the write operation of the TDC output ($binary\_out$) when $cim_{en}$ is enabled.}}
\label{fig:res_driver_sim}
\end{figure}

}
{\color{black}
\subsection{Weight Stationary Data Mapping Scheme}
\label{sec:weight_stationary}
%\vspace{-0.80cm}

\begin{figure}[t!]
\begin{center}
\includegraphics[width = 0.5\textwidth]{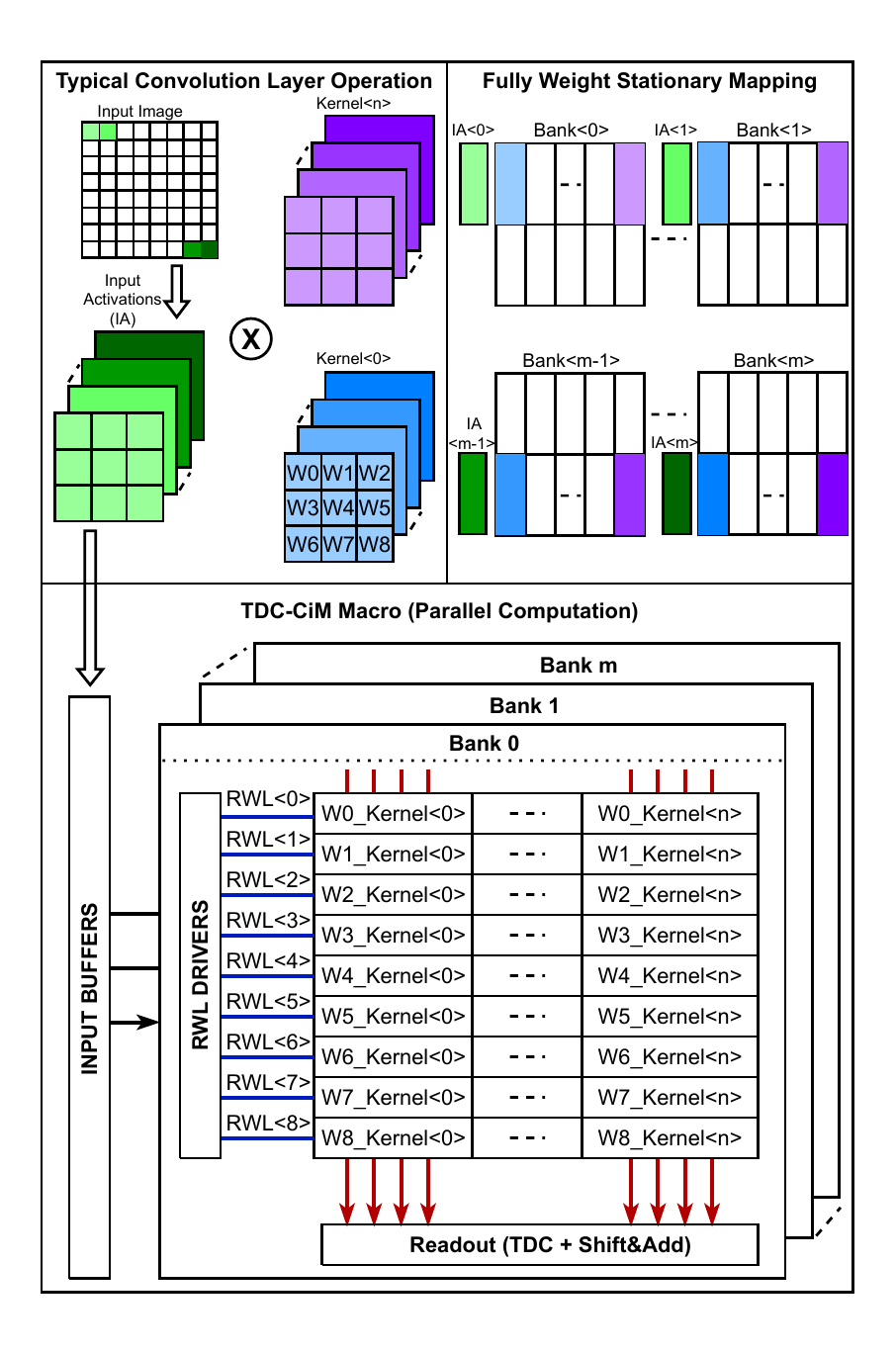}
\end{center}
\vspace{-0.40cm}
\caption{The fully row-stationary weight mapping scheme of the proposed TDC-CiM Macro only requires sending the input activations each computation cycle, significantly reducing the latency for off-chip access of weights.}
\label{fig:layer_mapping_ws}
\vspace{-0.60cm}
\end{figure}

Figure~\ref{fig:layer_mapping_ws} illustrates the fully weight stationary data mapping scheme employed in the proposed TDC-CiM macro for CNN workloads. In this scheme, the 8-bit kernel weights for each CNN layer are first loaded and held stationary inside the SRAM banks, requiring only a single initialization cycle. This configuration minimizes unnecessary data movement by keeping the kernel weights local to each bank, while the input feature maps (IFMs) are supplied dynamically through the row wordline (RWL) drivers during computation. Towards data mapping, each TDC-CiM row stores 8-bit weights, and the input buffer employs two rows to store the 4-bit MSBs and 4-bit LSBs of the IFM from different channels. The weight stationary arrangement provides additional flexibility in how IFMs are mapped across the banks. When the number of kernels is large, the same IFM can be broadcast to all banks in parallel, allowing each bank to process a different kernel set with the same input data. Conversely, when the kernel count is small, multiple copies of the IFM can be distributed across different banks, enabling parallel computation on multiple IFM tiles within the same cycle. This choice between kernel parallelism and input parallelism helps balance throughput and utilization based on layer requirements. After each matrix-vector multiplication, the output feature maps are written back into the banks, ready for the next layer. This mapping strategy significantly reduces off-chip weight transfers and improves energy efficiency by exploiting local data reuse.

For example, in a typical LeNet-5 layer, the number of convolutional kernels is relatively small compared to large modern CNNs. In this case, the proposed weight stationary scheme can map the limited kernel set across only a portion of the available SRAM banks, freeing the remaining banks to process multiple IFM tiles in parallel during the same computation cycle. This allows the architecture to maintain high utilization and throughput even when the kernel count is low, by distributing different portions of the input data across separate banks for concurrent processing. For layers with larger kernel sets, the same mapping strategy can instead broadcast a single IFM to all banks to process a higher number of kernels simultaneously. This flexibility ensures that the TDC-CiM macro can adapt to diverse CNN layers, while minimizing redundant memory access and maximizing parallel computation.
{\color{black}
The control circuitry of the proposed weight stationary mapping scheme includes a programmable row decoder that enables selective activation of multiple RWLs across banks based on IFM positions, and a column decoder capable of indexing stationary kernel weights stored within the bitcells. The main controller coordinates these components to enable synchronized broadcast of IFM slices across banks and sequential MAC operations. Additionally, cycle-level control logic is required to manage the multi-cycle processing of 8-bit IFMs using two 4-bit partial accumulations per MAC operation.}

\begin{algorithm}[h]
\renewcommand{\thealgorithm}{}
\renewcommand{\baselinestretch}{1}
\caption{\textbf{I}. Mapping Quantized CNN Workloads to Optimal TDC-CiM Architecture}
\label{alg1:cnn_recipe}
\begin{algorithmic}[1]

\State {\bf Input:} Pretrained Neural Network ($NN$), SRAM Macros ($SRAM_{list}$);%~\label{alg1:cnn_input}

\State {\bf Output:} rTD--CiM Architecture;%~\label{alg1:cnn_output}

\State $QNN \gets Quantize(NN, 8\text{-}bit)$; \Comment{Quantize the NN to 8-bit INT precision}~\label{alg1:cnn_quantize}

\State $LayerInfo, Weights \gets Extract(QNN)$; \Comment{Extract layer-wise information and weights}~\label{alg1:cnn_extract}

\State $SRAM_{size} \gets IdentifySRAM(LayerInfo, Weights)$; \Comment{Determine a set of SRAM's based on layer info and kernel counts}~\label{alg1:cnn_select_sram}

\ForAll{$SRAM$ in $SRAM_{size}$} \Comment{Loop through each SRAM Macro}~\label{alg1:metrics_for}
    \State $Metrics_{list}[SRAM] \gets Evaluate(LayerInfo, Weights,SRAM)$; \Comment{Evaluate power, latency, and energy for each SRAM Macro chosen}~\label{alg1:cnn_evaluate}
\EndFor~\label{alg1:metrics_for_end}

\State $BestSRAM \gets SortEnergy(Metrics_{list})$; \Comment{Determine lowest energy consuming AIG}~\label{alg1:final_optimal}

\State $L_{res} \gets CalculateInductor(BestSRAM)$; \Comment{Compute resonant inductor size for chosen SRAM}~\label{alg1:cnn_inductor}

\State \textbf{Output:} TDC-CiM Architecture $\gets$ $\{BestSRAM, L_{res}\}$; \Comment{Resulting TDC-CiM architecture for the CNN model}~\label{alg1:cnn_output_architecture}

\end{algorithmic}
\end{algorithm}

Algorithm.~I describes the workflow for selecting an optimal SRAM macro size to map a quantized CNN workload onto the proposed TDC-CiM architecture. The input to the algorithm is a pretrained neural network and a list of available SRAM macro configurations. The output is the resulting TDC-CiM hardware configuration tailored for the given model. First, the pretrained CNN is quantized to 8-bit integer precision, in Line.~\ref{alg1:cnn_quantize}, to enable efficient low-bit in-memory computation. Next, Line.~\ref{alg1:cnn_extract} extracts layer-wise
structural information and the corresponding weights from the
quantized model. Using this $LayerInfo~\&~Weights$, Line.~\ref{alg1:cnn_select_sram} identifies a subset of suitable SRAM macro sizes based on layer dimensions, parameter count, and the varying level of parallelism.  

For each chosen SRAM macro, the algorithm evaluates the power, latency, and energy performance in Lines.~\ref{alg1:metrics_for}–\ref{alg1:cnn_evaluate}, considering the specific workload and hardware constraints. Once the evaluations are complete, the configuration yielding the lowest total energy consumption is selected by the filtering step in Line.~\ref{alg1:final_optimal}. Finally, Line.~\ref{alg1:cnn_inductor} computes the required resonant inductor value for the selected SRAM size. This proposed methodology would result in the workload-specific TDC-CiM architecture that preserves quantization accuracy while maximizing energy efficiency and performance.

}
\section{Experimental Results}
\subsection{Experimental Setup}

An 8~KB SRAM memory instance is designed and simulated using 28~nm TSMC technology. The SRAM memory array comprises $256\times256$ 8T bitcells, implemented using Cadence Virtuoso, and simulations were performed utilizing the Cadence Spectre simulator at 0.5~GHz clock frequency. The inputs are driven using nine RWL drivers corresponding to a $3\times3$ input kernel of the $IFM$. The capacitor array, performing MAC computations, comprises binary-weighted bitwise capacitors and an $C_{acc}$. The bitwise LSB capacitor value is set to $1C = 4~\mathrm{fF}$, and the accumulation capacitor that stores the analog MAC output $V_{mac}$ is set to $32~\mathrm{fF}$. The readout circuitry comprises 64 TDC blocks, each connected to one capacitor array.

For system-level benchmarking, six neural network models were used, namely, LeNet-5~\cite{lenet} (MNIST~\cite{mnist}), MobileNetV1~\cite{MobileNetsv1} and MobileNetV2~\cite{MobileNetV2} (CIFAR-10~\cite{cifar}), SqueezeNet~\cite{squeezenet} and ResNet-18~\cite{resnet}(ImageNet-1K~\cite{imagenet}), and Tiny-YOLOv3~\cite{tinyyolov3} (COCO~\cite{coco}). All models were quantized to 8-bit precision using post-training quantization, without further fine-tuning. The proposed SRAM macro selection algorithm was applied to each network to identify the optimal SRAM size for minimizing energy consumption and inference latency. Energy and latency metrics were derived by mapping each layer to the SRAM configurations and accumulating per-layer estimates across the full network inference.

%{\TODO not claer} The output of the TDC blocks ($D_{out}$) is latched and provided as input data ($D_{in}$) to the resonant write drivers for writing the MAC output into the SRAM array.

%\subsection{MAC funtionality}

\subsection{TDC characterization and comparison}

\begin{figure}[b]
%\vspace{-0.65cm}
\begin{center}
\includegraphics[width = 0.4\textwidth]{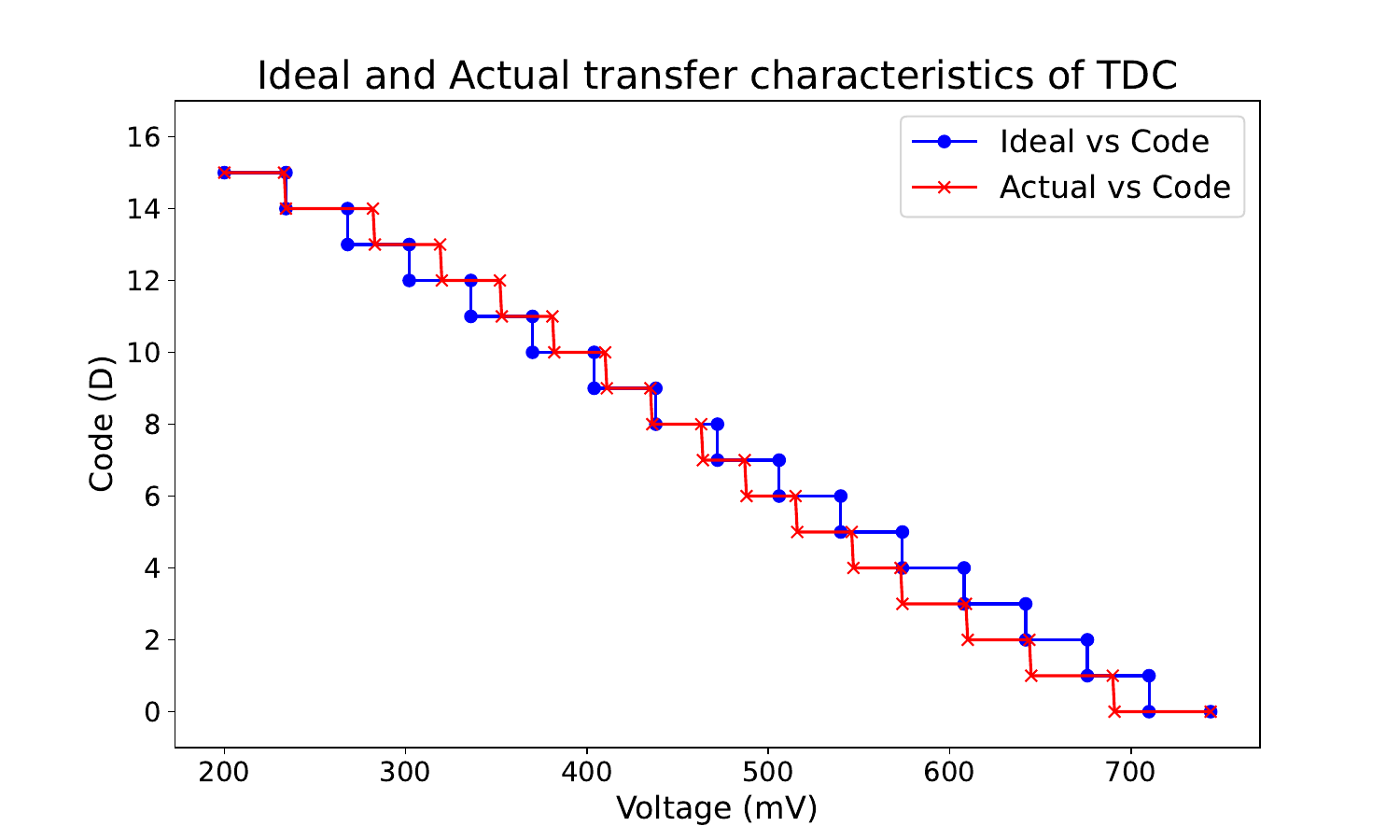}
\end{center}
%\caption{Functionality waveform illustrating NAND computation with "01"/"10" data and a successful energy-recycling writeback operation of the result.}
%%\vspace{-0.5cm}
\caption{{Simulated transfer characteristics of TDC show a deviation from the linear ideal characteristic curve but track the analog MAC output voltage $V_{mac}$.}}
\label{fig:tdc_characteristics}
%%\vspace{-0.3cm}
\end{figure}

\begin{figure}[b]
%%\vspace{-0.3cm}
\begin{center}
\includegraphics[width = 0.4\textwidth]{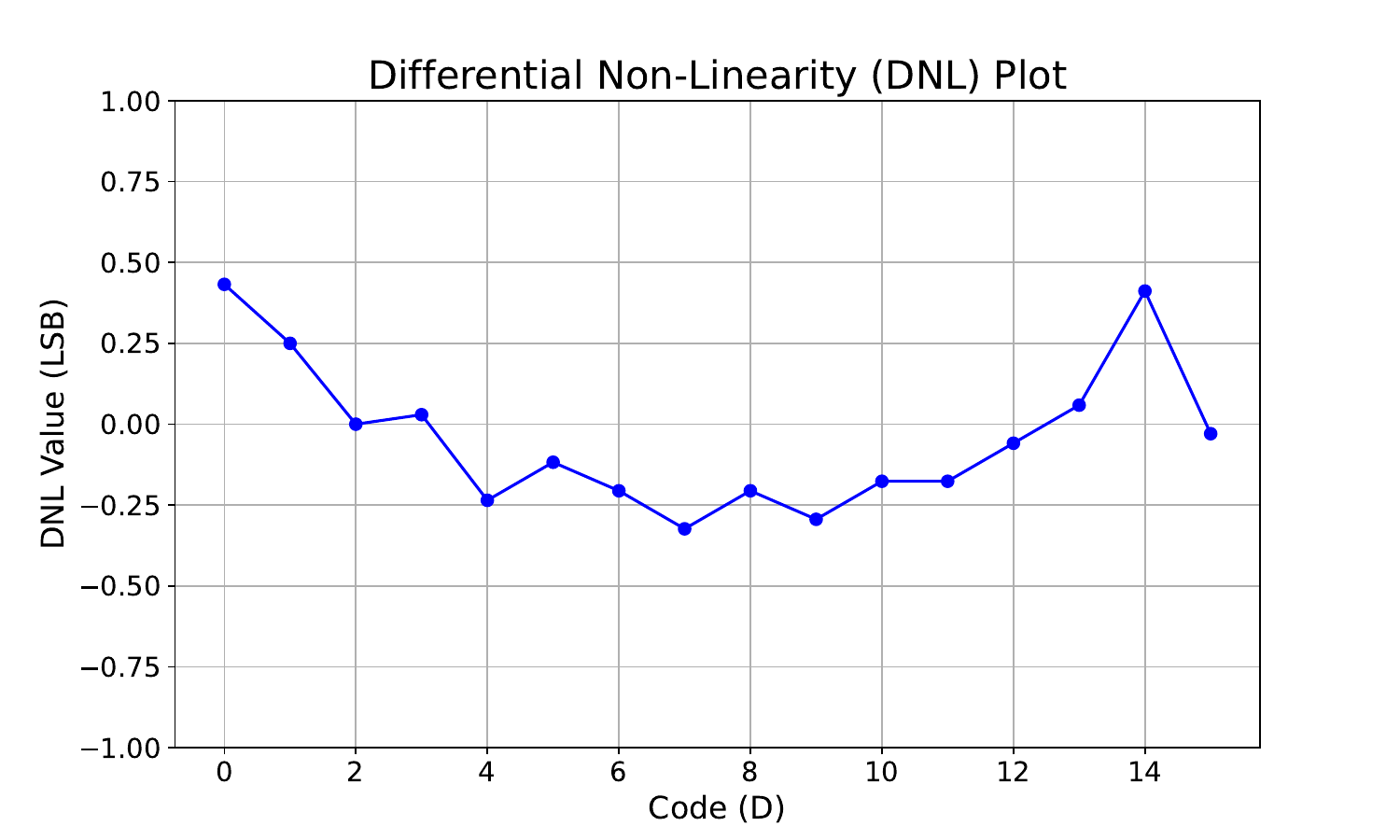}
\end{center}
%\caption{Functionality waveform illustrating NAND computation with "01"/"10" data and a successful energy-recycling writeback operation of the result.}
%%\vspace{-0.5cm}
\caption{{The DNL characteristics plot showcases a tolerable deviation of -0.3/+0.4 LSBs.}}
\label{fig:dnl_plot}
%%\vspace{-0.30cm}
\end{figure}

\begin{figure}[t]
%%\vspace{-0.5cm}
\begin{center}
\includegraphics[width = 0.4\textwidth]{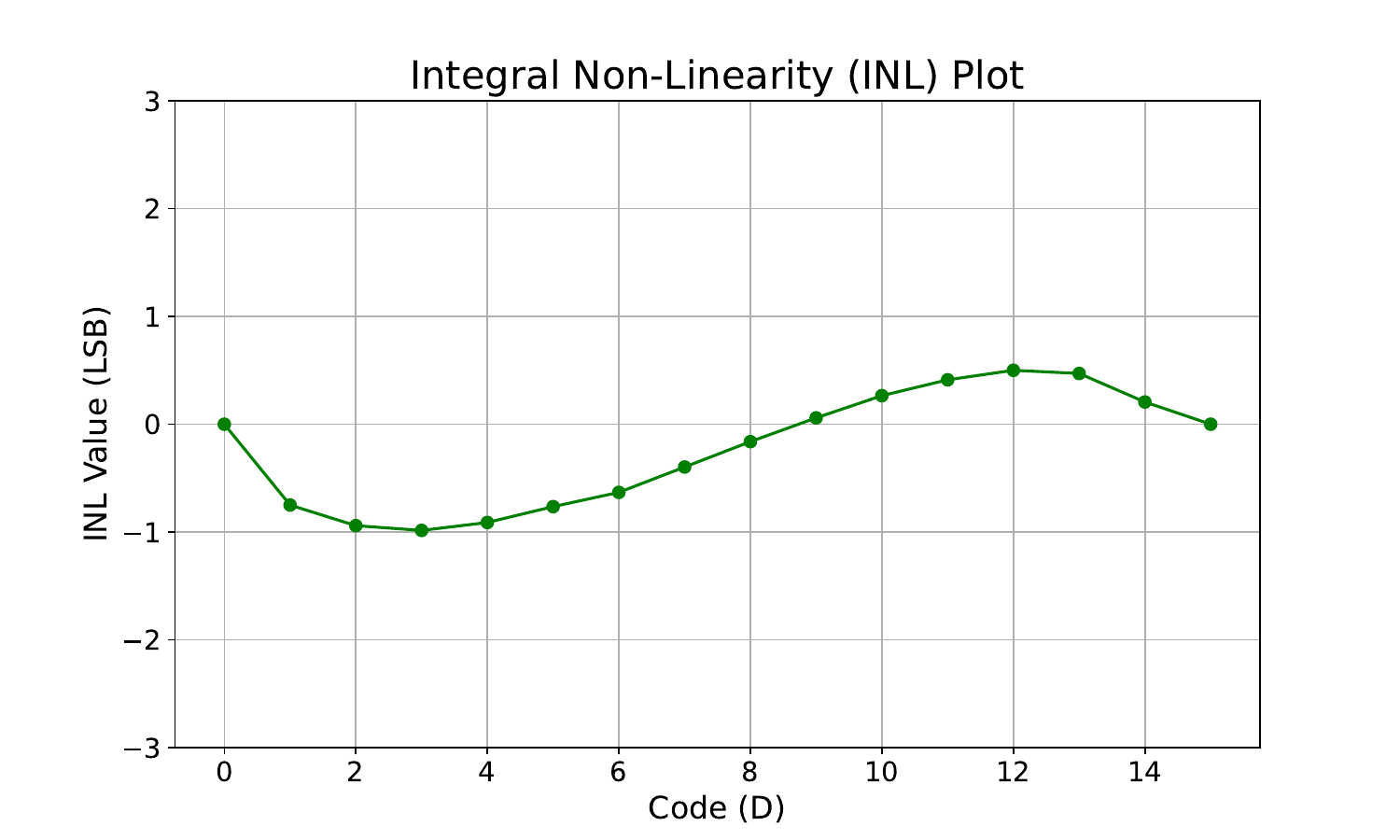}
\end{center}
%\caption{Functionality waveform illustrating NAND computation with "01"/"10" data and a successful energy-recycling writeback operation of the result.}
%%\vspace{-0.5cm}
\caption{{The INL characteristics plot of TDC demonstrates a minimum deviation of -1 LSBs for digital code ``2'' and a maximum deviation of +0.5 LSBs for ``12.''}}
\label{fig:inl_plot}
%%\vspace{-0.30cm}
\end{figure}

Figure.~\ref{fig:tdc_characteristics} illustrates the transfer characteristics to analyze the linearity of the TDC. The input voltage applied to the TDC is plotted on the x-axis, with a baseline offset of 200~$mV$. The full voltage range of the proposed TDC circuit is from 200~$mV$ to 800~$mV$. %On the y-axis, the resultant digital output, denoted as ``Code (D),'' is mapped in the range of 0 to 15.
The ideal performance of the TDC is depicted by the blue line, characterized by a consistent, monotonically decreasing function. In contrast, the simulated actual transfer characteristics of the proposed TDC circuit, shown in red, deviate from the linear function. This characteristic aligns with the linear discharge observed on the $C_{acc}$ as shown in Figure~\ref{fig:caps_discharge}. The offset of the actual transfer curve can be corrected by performing the Differential Non-Linearity (DNL) and Integral Non-Linearity (INL) analysis, shown in Figure~\ref{fig:dnl_plot} and Figure~\ref{fig:inl_plot}, respectively.

Figure.~\ref{fig:dnl_plot} illustrates the DNL characteristics of the proposed TDC. The DNL is a measure of the deviation from the ideal step size between consecutive codes expressed in Least Significant Bits (LSB). %The x-axis shows the code (D), which represents the digital output, and the y-axis shows the DNL values in LSBs. 
A positive DNL value indicates that the actual step size between two successive codes is larger than one LSB, while a negative DNL value means the step size is smaller than one LSB. If the DNL value ever reaches -1 LSB, it would indicate a missing code, which is a serious non-linearity error in the TDC. However, the graph indicates that all of the DNL values are within $\pm0.5$ LSB, which is considered to be tolerable bounds for TDC characterization. The plot showcases a maximum DNL value of +0.4 LSB for digital code ``0,'' and a minimum DNL value of -0.3 LSB for digital code ``7.''

Figure.~\ref{fig:inl_plot} shows the INL plot for the proposed TDC. INL is a measure of the converter's linearity, indicating the maximum deviation from the ideal function mapping of input to output over the full range of the converter. %The x-axis represents the digital output codes (D) produced by the TDC, and the y-axis shows the INL values in LSBs. 
The INL plot exhibits an initial positive deviation, signifying that the early output codes from the TDC are larger than the expected values for an ideally linear system. As the input value increases, the INL plot trends downward, eventually falling below the zero level, which indicates that the output codes from the TDC are incrementally lower than what would be predicted by a linear model. The plot shows a maximum INL value of +0.5 LSB for digital code ``12,'' and a minimum INL value of -1 LSB for digital code ``2.''

\begin{table*}[t]
    \centering
    \caption{Comparison of the TDC architecture with prior TD and SAR ADCs show lower power consumption of 45.6\% compared to~\cite{tdc_comp_1}, and 96\% compared to~\cite{tdc_comp_2}.}
    %%\vspace{-0.4cm}
    \label{tab:tdc_comparison}
    \resizebox{\linewidth}{!}{
    \begin{tabular}{|l|c|c|c|c|c|}
    \hline
    \textbf{Reference} & APCCAS'22~\cite{tdc_comp_1} & ESSCIRC'23\cite{tdc_comp_2} & RFIC'18\cite{tdc_comp_3} & JSSC'19~\cite{tdc_comp_4} & \textbf{This Work} \\ 
    \hline
    \textbf{Architecture}  & TD ADC & SAR ADC & SAR ADC & TD ADC & \textbf{TDC}\\
    \hline
    \textbf{Technology (nm})& 28 & 28 & 28 & 65 & \textbf{28} \\
    \hline
    \textbf{Resolution (bits)}& 9 & 6 & 8 & 8 & \textbf{4} \\
    \hline
    \textbf{Fs (GS/s)} & 0.5 & 1.4 & 8.8 & 1 & \textbf{1} \\
    \hline
    \textbf{Supply (V)} & 0.9 & 0.9 & 1.5 & 1 & \textbf{1} \\
    \hline
    \textbf{SNDR (dB)} & 54.69 & 67 & 38.4 & 45 & \textbf{19.45}\\
    \hline
    \textbf{SFDR (dB)} & 55.16 & NR & 48.9 & 60.3 & \textbf{22.4}\\
    \hline
    \textbf{Power (mW)} & 4.27 & 32 & 83.4 & 2.3 & \textbf{1.25} \\
    \hline
   \textbf{FoM (fJ/conv.step)} & 19.29 & 143.2 & 139.5 & 18.7 & \textbf{162.8} \\
    \hline
    \end{tabular}
    }
%%\vspace{-0.4cm}
\end{table*}

Table.~\ref{tab:tdc_comparison} compares various Analog-to-Digital Converter (ADC) architectures across several design references. The resolution of the TDC used in this work is 4 bits at a sampling rate of 1GS/s. The voltage supply is 1V for the TDC. The TDC exhibits an SNDR of 19.45 dB and an SFDR of 22.4 dB. The TDC utilizes $V_{mac}$ as its input, eliminating the need for the voltage-to-time converters typically required in conventional TDC architectures. This approach reduces the overall power consumption of the TDC framework. The proposed TDC achieves 71\% lower power consumption than \cite{tdc_comp_1} and 45.6\% lower power consumption than~\cite{tdc_comp_4}, which are time-domain analog-to-digital converters ($TD\ ADC$). Additionally, the proposed TDC demonstrates 98\% lower power consumption than~\cite{tdc_comp_3} and 96\% lower power consumption than~\cite{tdc_comp_2}, which are $SAR\ ADCs$, typically employed in conventional CIM architectures. The TDC achieves a Walden Figure of Merit~($FoM$) of 162.8 fJ/conversion step, which is 12\% higher than~\cite{tdc_comp_2} and 14\% higher than~\cite{tdc_comp_3}. This FoM, which is directly correlated to the bit resolution,  can be reduced by increasing the number of output bits of the TDC.

%{

\subsection{Process, Voltage, and Temperature Variation Analysis}

% is this paragraph reapeated in next paragraphs?
%To evaluate the robustness of the proposed TDC design, we performed extensive Monte Carlo simulations considering process, voltage, and temperature (PVT) variations. A total of 2000 runs were conducted using $3\sigma$ deviations. The supply voltage is varied by $\pm10\%$ from its nominal value to assess the impact of voltage scaling, while the operating temperature was swept across three representative corners: $0\,^{\circ}\mathrm{C}$, $27\,^{\circ}\mathrm{C}$, and $100\,^{\circ}\mathrm{C}$. This comprehensive PVT analysis ensures that the TDC maintains stable performance and acceptable variation margins under practical operating conditions.

\begin{figure*}[h]
%%\vspace{-0.5cm}
\begin{center}
\includegraphics[width = \textwidth]{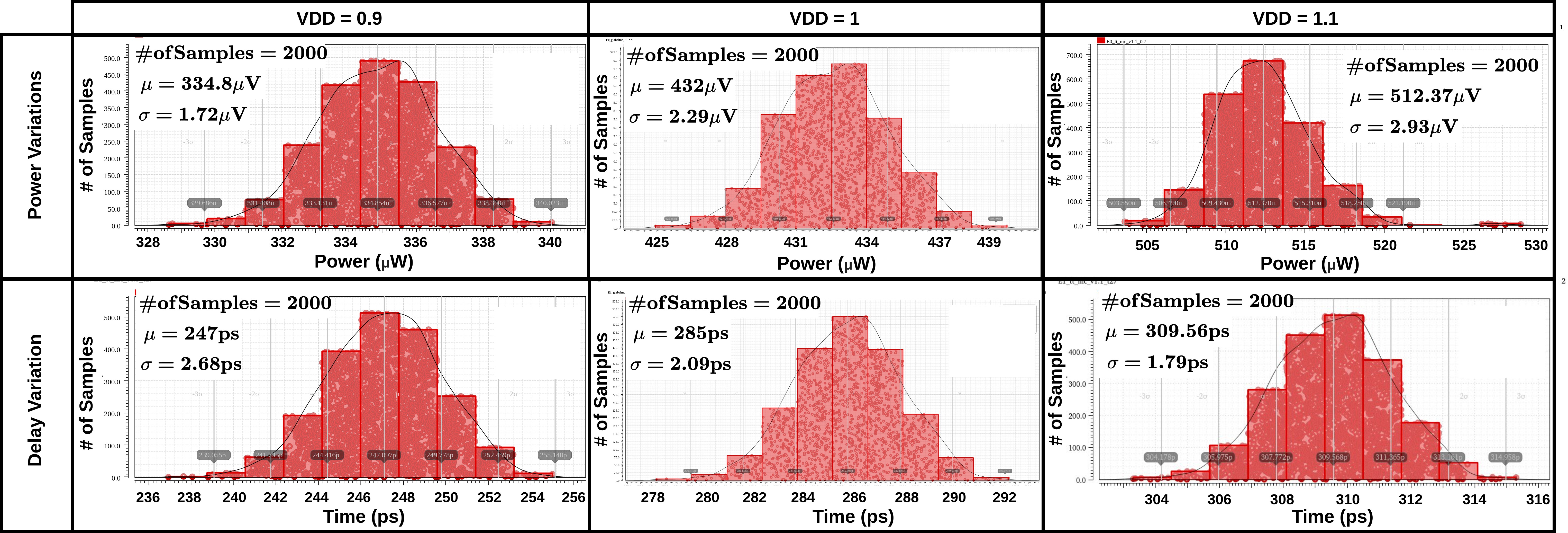}
\end{center}
%\caption{Functionality waveform illustrating NAND computation with "01"/"10" data and a successful energy-recycling writeback operation of the result.}
%%\vspace{-0.5cm}
\caption{Monte Carlo analysis of the proposed TDC under supply voltage variations. The top row shows the power variation for $V_{\mathrm{DD}} = 0.9\,\mathrm{V}$, $1\,\mathrm{V}$, and $1.1\,\mathrm{V}$, respectively. The bottom row shows the corresponding delay variation distributions for each voltage corner. A total of 2000 samples were simulated for each case using $3\sigma$ process variations and $\pm10\%$ supply voltage variation, demonstrating that increasing the supply voltage increases the mean power consumption while reducing the mean delay, consistent with the expected behavior of voltage scaling in delay stages.}

\label{fig:voltage_variation_mc}
%%\vspace{-0.30cm}
\end{figure*}

To evaluate the robustness of the proposed TDC design, we performed extensive Monte Carlo simulations considering PVT variations~\cite{Islam_dcmcs:2018}.
We performed a Monte Carlo analysis of the proposed TDC under supply voltage variations, using 2000 samples for each voltage corner with a $\pm10\%$ variation from the nominal voltage. The results, shown in Fig.~\ref{fig:voltage_variation_mc}, illustrate the impact of supply voltage on power and delay characteristics. For $V_{\mathrm{DD}} = 0.9\,\mathrm{V}$, the mean power consumption is $334.8~\mu\mathrm{W}$ with a standard deviation of $1.72~\mu\mathrm{W}$, and the mean delay is $247~\mathrm{ps}$ with a standard deviation of $2.68~\mathrm{ps}$. Increasing the supply voltage to $V_{\mathrm{DD}} = 1\,\mathrm{V}$ results in a mean power of $432~\mu\mathrm{W}$ ($\sigma = 2.29~\mu\mathrm{W}$) and a mean delay of $285~\mathrm{ps}$ ($\sigma = 2.09~\mathrm{ps}$). At $V_{\mathrm{DD}} = 1.1\,\mathrm{V}$, the mean power further increases to $512.37~\mu\mathrm{W}$ with a standard deviation of $2.93~\mu\mathrm{W}$, while the mean delay reduces to $309.56~\mathrm{ps}$ with a standard deviation of $1.79~\mathrm{ps}$. The mean of the delay variation corresponds to the pulse width of a single delay stage, where a larger pulse width indicates a lower overall TDC delay. These results confirm that higher supply voltages increase power consumption but reduce delay, demonstrating the expected trade-off for voltage scaling in the TDC design.

\begin{figure*}[h]
%%\vspace{-0.5cm}
\begin{center}
\includegraphics[width = \textwidth]{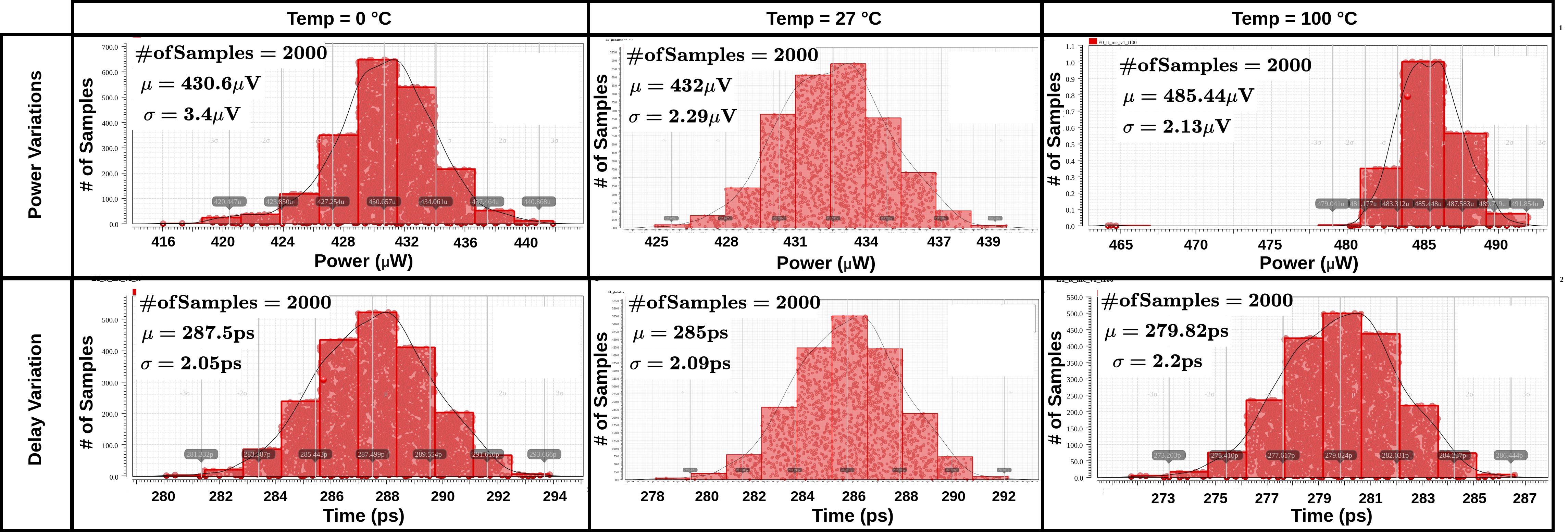}
\end{center}
%\caption{Functionality waveform illustrating NAND computation with "01"/"10" data and a successful energy-recycling writeback operation of the result.}
%%\vspace{-0.5cm}
\caption{Monte Carlo analysis of the proposed TDC under temperature variations. The top row shows power variation for $0\,^{\circ}\mathrm{C}$, $27\,^{\circ}\mathrm{C}$, and $100\,^{\circ}\mathrm{C}$, while the bottom row shows the corresponding delay variation. Each corner uses 2000 samples with $3\sigma$ process variation, showing that higher temperatures slightly increase power and reduce delay while maintaining stable performance across the full operating range.}

\label{fig:temperature_variation_mc}
%%\vspace{-0.30cm}
\end{figure*}

We also performed a Monte Carlo analysis of the proposed TDC under temperature variations, using 2000 samples for each temperature corner. The temperature was varied across three representative operating points: $0\,^{\circ}\mathrm{C}$, $27\,^{\circ}\mathrm{C}$, and $100\,^{\circ}\mathrm{C}$, to assess the impact of thermal conditions on power and delay performance. As shown in Fig.~\ref{fig:temperature_variation_mc}, the mean power consumption at $0\,^{\circ}\mathrm{C}$ is $430.6~\mu\mathrm{W}$ with a standard deviation of $3.4~\mu\mathrm{W}$, while the mean delay is $287.5~\mathrm{ps}$ with a standard deviation of $2.05~\mathrm{ps}$. At the nominal temperature of $27\,^{\circ}\mathrm{C}$, the mean power is $432~\mu\mathrm{W}$ with $\sigma = 2.29~\mu\mathrm{W}$ and the mean delay is $285~\mathrm{ps}$ with $\sigma = 2.09~\mathrm{ps}$. For high temperature operation at $100\,^{\circ}\mathrm{C}$, the mean power increases to $485.44~\mu\mathrm{W}$ with a standard deviation of $2.13~\mu\mathrm{W}$, and the mean delay decreases to $279.82~\mathrm{ps}$ with a standard deviation of $2.2~\mathrm{ps}$. Similar to the voltage variation results, the mean of the delay variation represents the pulse width of a single delay stage, where a higher pulse width corresponds to a lower overall TDC delay. These results indicate that higher temperatures slightly increase power consumption and reduce delay due to enhanced carrier mobility at elevated temperatures.

\subsection{Benchmarking with Neural Network Models}

We have evaluated the proposed TDC-CiM architecture on six different neural network models, ranging from small networks with 60K parameters to large models with over 11M parameters. Specifically, we used the LeNet-5~\cite{lenet} model trained on the MNIST~\cite{mnist} dataset, MobileNetV1~\cite{MobileNetsv1} and MobileNetV2~\cite{MobileNetV2} trained on the CIFAR-10~\cite{cifar} dataset, SqueezeNet~\cite{squeezenet} and ResNet trained on the ImageNet-1K~\cite{imagenet} dataset, and Tiny-YOLOv3~\cite{tinyyolov3} trained on the COCO~\cite{coco} dataset. 

\begin{figure*}[t]
%%\vspace{-0.5cm}
\begin{center}
\includegraphics[width = \textwidth]{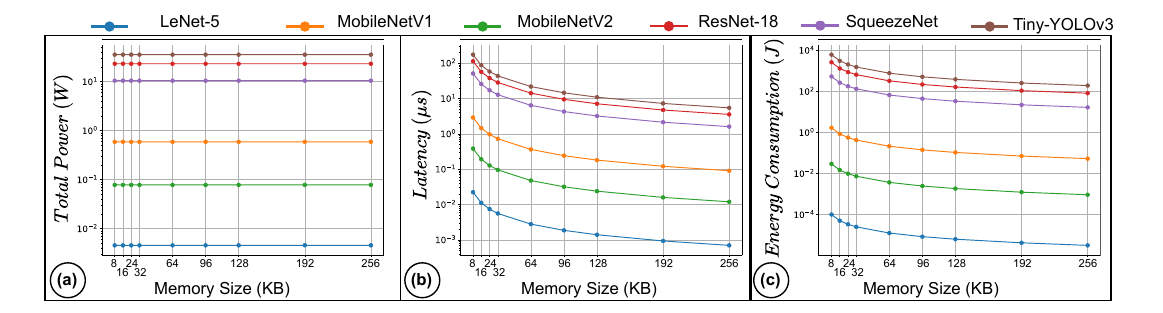}
\end{center}
%\caption{Functionality waveform illustrating NAND computation with "01"/"10" data and a successful energy-recycling writeback operation of the result.}
%%\vspace{-0.5cm}
\caption{Evaluation of CNN models on SRAM macros ranging from 8~KB to 256~KB. (a) Total power consumption remains mostly constant across memory sizes, with deeper models consuming more power. (b) Inference latency decreases with larger SRAM sizes, with greater reduction observed in deeper models due to fewer memory accesses. (c) Energy consumption decreases with memory size, indicating improved efficiency from reduced off-chip data movement.}
\label{fig:benchmarks_nn}
%%\vspace{-0.30cm}
\end{figure*}

Table~\ref{tab:benchmark_acc} reports the full-precision accuracy alongside the quantized INT8 accuracy, demonstrating that quantization introduces minimal accuracy loss. The pretrained models were quantized without any further fine-tuning. Depending on the model size, varying SRAM sizes enable more energy-efficient implementations. Figure~\ref{fig:benchmarks_nn} shows the power, latency, and energy consumption comparison for these neural network benchmarks. A detailed power, latency, and energy analysis was performed across nine different SRAM macro sizes and varying numbers of banks.

Figure~\ref{fig:benchmarks_nn}(a) compares the overall power consumption of each neural network benchmark when executed on the proposed TDC-CiM architecture across varying SRAM macro sizes. The total power consumption remains approximately constant for a given workload, regardless of the SRAM size. This behavior can be explained by considering the balance between parallelism and cycle count: for instance, a 16~KB SRAM macro provides twice the storage compared to an 8~KB macro, allowing twice as many MAC operations to be computed in parallel per clock cycle. However, the smaller 8~KB configuration requires twice as many cycles to complete the same total number of MAC operations for a fixed model size. As a result, the total dynamic switching activity remains nearly unchanged, and the static power overhead is effectively spread across the entire computation time, leading to negligible variation in overall power consumption across different SRAM macro sizes.

Figure~\ref{fig:benchmarks_nn}(b) depicts the inference latency for all neural network benchmarks across different SRAM macro sizes. On average, there is a latency reduction of ~70\% when increasing the macro size from 8~KB to 24~KB, highlighting the benefit of greater local storage in enabling more MAC operations per cycle and reducing the number of sequential compute cycles. A further increase from 16~KB to 64~KB achieves an additional average latency reduction of around 60\%, demonstrating the improved parallel data access and reduced need for repeated weight fetching. Similarly, scaling the macro from 128~KB to 256~KB yields an average latency improvement of about 45\%, which shows diminishing returns at larger macro sizes due to saturation of available parallelism relative to the workload size. This clear trend confirms that larger SRAM macros in the TDC-CiM architecture effectively exploit intra-cycle parallelism and data reuse, minimizing memory bottlenecks and accelerating inference performance for a wide range of network sizes.

Figure~\ref{fig:benchmarks_nn}(c) illustrates the average energy consumption per inference for all benchmark models across different SRAM macro sizes. On average, expanding the macro from 8~KB to 32~KB reduces the total energy by 67\%, primarily by minimizing the number of sequential memory access cycles required for weight and activation fetches. Increasing the macro size further from 32~KB to 96~KB yields an additional average energy reduction of 73.6\%, highlighting the advantage of higher data reuse and fewer off-chip transactions. Extending the SRAM from 96~KB to 256~KB results in a further 62\% drop in energy, showing that larger macros maintain significant efficiency gains even at high SRAM memory size. These results confirm that the proposed TDC-CiM architecture can achieve substantial energy savings by leveraging large SRAM arrays to maximize in-memory operations and reduce redundant data movement between compute and storage units.

\begin{table*}[t!]
\centering
\caption{{Benchmarking analysis of the neural network models reveals a minimal accuracy drop, considering both full precision and INT8 data types, as well as the optimal SRAM sizes for achieving the optimal energy efficiency.}}
\label{tab:benchmark_acc}
\renewcommand{\arraystretch}{1.8} % Adjust row height
\resizebox{\linewidth}{!}{
{\large
\begin{tabular}{|c|c|c|c|c|c|c|}
\hline
\textbf{Model} & \textbf{Dataset} & \textbf{Parameters} & \textbf{Full Precision Accuracy} & \textbf{\shortstack[c]{INT8 Accuracy\\(Software)}} & \textbf{\shortstack[c]{INT8 Accuracy\\(TDC-CiM)}} & \textbf{SRAM Size} \\ \hline
\textbf{LeNet-5~\cite{lenet}} & MNIST~\cite{mnist} & 61.47K & 98.76\% & 98.7\% & 98.7\%  & 24 KB\\ \hline
\textbf{MobileNetV1~\cite{MobileNetsv1}} & CIFAR-10~\cite{cifar} & 3.19M & 92.45\% & 89.73\% & 89.6\%& 64 KB\\ \hline
\textbf{MobileNetV2~\cite{MobileNetV2}} & CIFAR-10~\cite{cifar} & 2.35M & 96.43\% & 96.48\% & 96.1\% & 64 KB\\ \hline
\textbf{SqueezeNet~\cite{squeezenet}} & ImageNet-1K~\cite{imagenet} & 1.24M & 58.18\% &56.50\%& 55.80\% &  128 KB\\ \hline
\textbf{ResNet-18~\cite{resnet}} & ImageNet-1K~\cite{imagenet} & 11.68M & 67.61\% &67.14\% & 67.1\%  & 256 KB\\ \hline
\textbf{Tiny-YOLOv3~\cite{tinyyolov3}} & COCO~\cite{coco} & 8.85M & 33\% (mAP) & 31\% (mAP) & 31\% (mAP) & 256 KB\\ \hline
\end{tabular}}%
}

\end{table*}

Table~\ref{tab:benchmark_acc} summarizes the benchmarked neural network models, including dataset, parameter count, accuracy before and after INT8 quantization, and the selected SRAM macro size used for optimal inference performance. On average, the quantization to INT8 results in an average accuracy reduction of 0.75\% across the five CNN models evaluated. The optimal SRAM size for each model balances memory capacity with the required level of parallel computation. Smaller models such as LeNet-5~\cite{lenet} can achieve maximum efficiency with relatively small 24~KB SRAM blocks since their total parameter count is low. In contrast, larger models like ResNet-18~\cite{resnet} and Tiny-YOLOv3~\cite{tinyyolov3} require significantly more on-chip storage to exploit high levels of parallel MAC operations and minimize repeated weight fetching, justifying the use of large 256~KB SRAM configurations. Although SqueezeNet~\cite{squeezenet} has a relatively low parameter count compared to other ImageNet-class models, its architectural design involves a large number of operations per parameter due to its squeeze-and-expand modules. As a result, it demands a higher 128~KB SRAM size than its raw parameter count might suggest, to sustain high parallelism and reduce the overall cycle count. Table~\ref{tab:benchmark_acc} also demonstrates that selecting the appropriate SRAM size for each network ensures low quantization-induced accuracy drop while achieving substantial energy and latency improvements through compute-in-memory parallelism.

%{
\subsection{MAC comparison with previous works}
%%%\vspace{-1 pt}
{
Table~\ref{tab:mac_comparison} compares the proposed TDC-CiM architecture with existing CiM architectures capable of performing MAC operations. The proposed architecture can perform MAC operations for 8-bit inputs and 8-bit weights using 8T bitcells at 0.5~$GHz$ clock frequency. Table~\ref{tab:mac_comparison} reports our numbers from transistor-level, post-layout SPICE simulations of \(512\times256\) 8T TDC-CIM macro that includes wordline and bitline drivers, local control, and the 4-bit TDC. All simulations use a 28 nm CMOS process at TT and \(25\,^\circ\mathrm{C}\) with \(VDD=1.0\,\mathrm{V}\). The TDC input range is \(200\) to \(800\,\mathrm{mV}\) based on the MAC discharge range. INT8 inference is implemented as two 4-bit MAC cycles with shift and add accumulation. Throughput is computed from simulated cycle counts and clock frequency using the configured parallelism for a 16 KB SRAM macro. Energy efficiency(TOPS/W) is the throughput divided by total macro power, where the power includes dynamic and leakage from the CiM array, the wordline and bitline drivers, the TDC readout, and local control.}

\begin{table}[t]
    \centering
    %%\vspace{-0.2cm}
    \captionsetup{justification=centering} % Centering the caption
    \caption{{Comparison of the proposed TDC-CiM architecture with prior CiM works showcases 59.7\% higher energy efficiency compared to~\cite{comp_jssc_23} and achieves $6.25\times$ higher throughput than~\cite{comp_jssc_24}.}}
    %\vspace{-0.4cm}
    \label{tab:mac_comparison}
    \large
    \centering
    \resizebox{\linewidth}{!}{%
        \begin{tabular}{|l|c|c|c|c|c|}
            \hline
            & \textbf{This work} & JSSC'23~\cite{comp_jssc_23} & JSSC'24~\cite{comp_jssc_24} & TCAS-I'24~\cite{comp_tcasi_24} & ISSCC'22\cite{comp_isscc_22} \\
            \hline
            \textbf{Technology (nm)} & \textbf{28} & 22 & 28 & 28 & 28 \\
            \hline
            \textbf{Cell Type} & \textbf{8T} & 9T & 8T & 8T1C & 6T \\
            \hline
            \textbf{Array Size} & \textbf{16KB} & 128KB & 16KB & 64KB & 1MB \\
            \hline
            \textbf{Precision (input/weight)} & \textbf{8/8} & 8/8 & 8/8 & 8/8 & 8/8 \\
            \hline
            \textbf{Supply Voltage (V)} & \textbf{1} & 1.1 & 0.9 & 0.8 & 0.9 \\
            \hline
            \textbf{Frequency (GHz)} & \textbf{0.5} & 0.24 & 0.2 & 0.25 & - \\
            \hline
            \textbf{Throughput (GOPS)} & \textbf{320} & 1000 & 51.2 & 256 & 1241 \\
            \hline
            \textbf{Energy Efficiency (TOPS/W)} & \textbf{38.46} & 15.5 & 22.2 & 71.17 & 37.01 \\
            \hline
            \textbf{Compute Density (TOPS/$mm^2$)} & \textbf{2.1} & 4 & 1.6 & 1.29 & - \\
             \hline
        \end{tabular}%
    }
    %\vspace{-0.65cm}
\end{table}

The TDC-CiM achieves a throughput of 320 $GOPS$ with an energy efficiency of 38.46 $TOPS/W$. In~\cite{comp_jssc_23}, a 9T bipolar cell with differential inputs is used to perform XOR operations, which leads to higher leakage power due to the continuous current paths. In contrast, the 8T cell employed in this work disconnects the capacitors when idle, effectively reducing leakage. As a result, the proposed architecture achieves twice the operating frequency and a 59.7\% improvement in energy efficiency compared to~\cite{comp_jssc_23}. In~\cite{comp_jssc_24}, DAC/ADC-based in-memory operations contribute to high energy consumption, while more than 50\% of the total area is occupied by the data converters used for CiM computations in~\cite{comp_tcasi_24}. In contrast, the proposed TDC-CiM architecture eliminates the need for an ADC, significantly reducing energy overhead and area usage. As a result, the TDC-CiM achieves a $6.25\times$ increase in throughput, $1.31\times$ higher area efficiency and a 42.2\% improvement in energy efficiency compared to~\cite{comp_jssc_24}, while also delivering $2\times$ higher frequency and 20\% higher throughput, and $1.62\times$ higher area efficiency than~\cite{comp_tcasi_24}. 

%}
\section{Conclusion}
This paper presents an TDC-CiM architecture designed for low-power in-memory computation. The architecture performs bitwise multiplications using 8T SRAM bitcells and leverages a binary-weighted capacitor array to execute MAC operations directly within the memory. A custom time-to-digital converter (TDC) digitizes the analog MAC results, eliminating the need for energy- and area-intensive ADCs commonly used in conventional CiM designs. The proposed TDC achieves a sampling frequency of 1~GS/s with 1.25~mW power consumption, an SNDR of 19.45~dB, and a Walden FoM of 162.8~fJ/conversion-step.

To support scalable deployment across CNN workloads, an automated SRAM macro selection algorithm is proposed using a fully weight stationary mapping scheme. Benchmarking across six neural networks with SRAM sizes ranging from 8~KB to 256~KB shows that increasing the SRAM size results in significant energy savings: up to 67\% from 8~KB to 32~KB, an additional 73.6\% from 32~KB to 96~KB, and a further 62\% from 96~KB to 256~KB. These improvements are attributed to reduced memory access cycles, increased in-memory data reuse, and improved MAC parallelism. 

The proposed TDC-CiM architecture demonstrates a throughput of 320~GOPS with an energy efficiency of 38.46~TOPS/W, confirming its potential for energy-constrained edge AI applications requiring parallel and low-latency MAC operations.

\bibliographystyle{IEEEtran}
\bibliography{main}

\begin{IEEEbiography}
[{\includegraphics[width=1in,height=1in,clip,keepaspectratio]{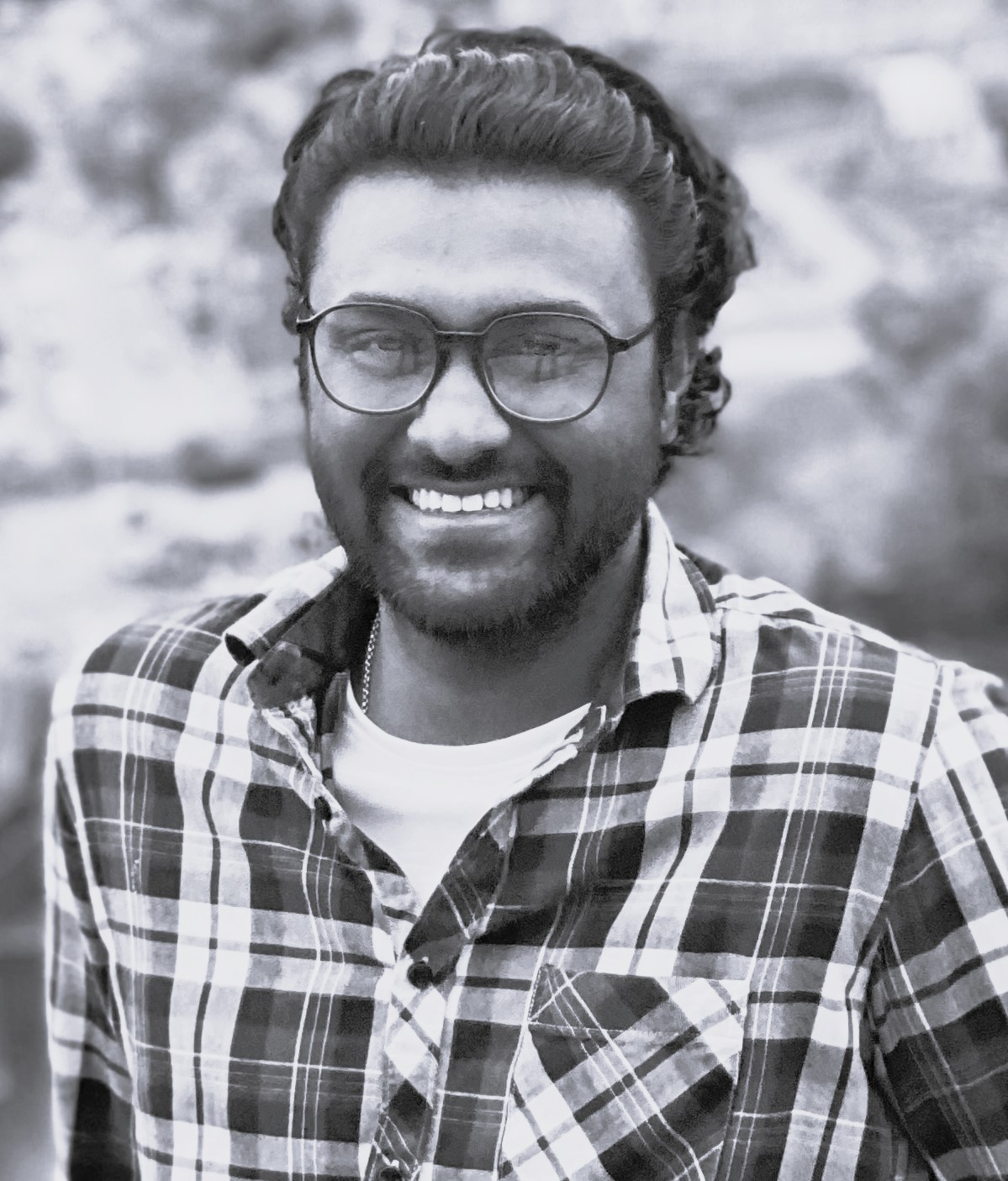}}] 
{Dhandeep Challagundla} (Student Member, IEEE) received his M.S degree from The University of Maryland Baltimore County (UMBC), MD, USA, where he is currently pursuing the Ph.D. degree with Computer Science and Electrical Engineering Department. His research interests revolve around energy-efficient computing, Compute-in-Memories, SRAM design, low-power circuit design, Mixed-signal IC design, and EDA tools.
\end{IEEEbiography}
%\vspace{-2.00cm}

\begin{IEEEbiography}
[{\includegraphics[width=1in,height=1.25in,clip,keepaspectratio]{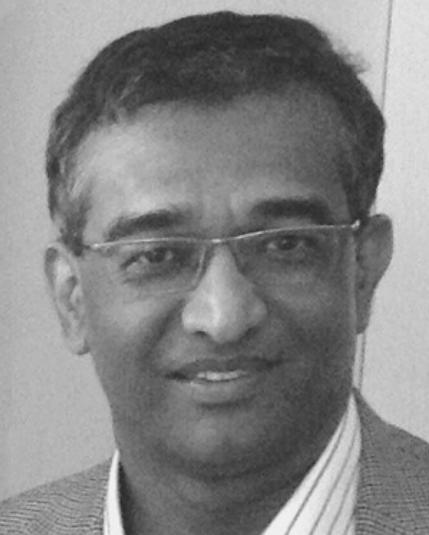}}] 
{Prof. Ignatius Bezzam} is a PhD graduate in Electrical Engineering from Santa Clara University (2015) and a Bachelor of Technology graduate of IIT Madras, India in 1983. Dr. Bezzam holds several key patents in Analog Mixed Signal Integrated Circuit (IC) design with publications in top international conferences, including the ISSCC, ESSCIRC and TCAS. Dr. Bezzam has owned 30 first silicon successes with global teams, with 33 years of next generation chip design experience in Silicon Valley, Europe and Asia.
\end{IEEEbiography}
%\vspace{-1.50cm}
\begin{IEEEbiography}[{\includegraphics[width=1in,height=1.25in,clip,keepaspectratio]{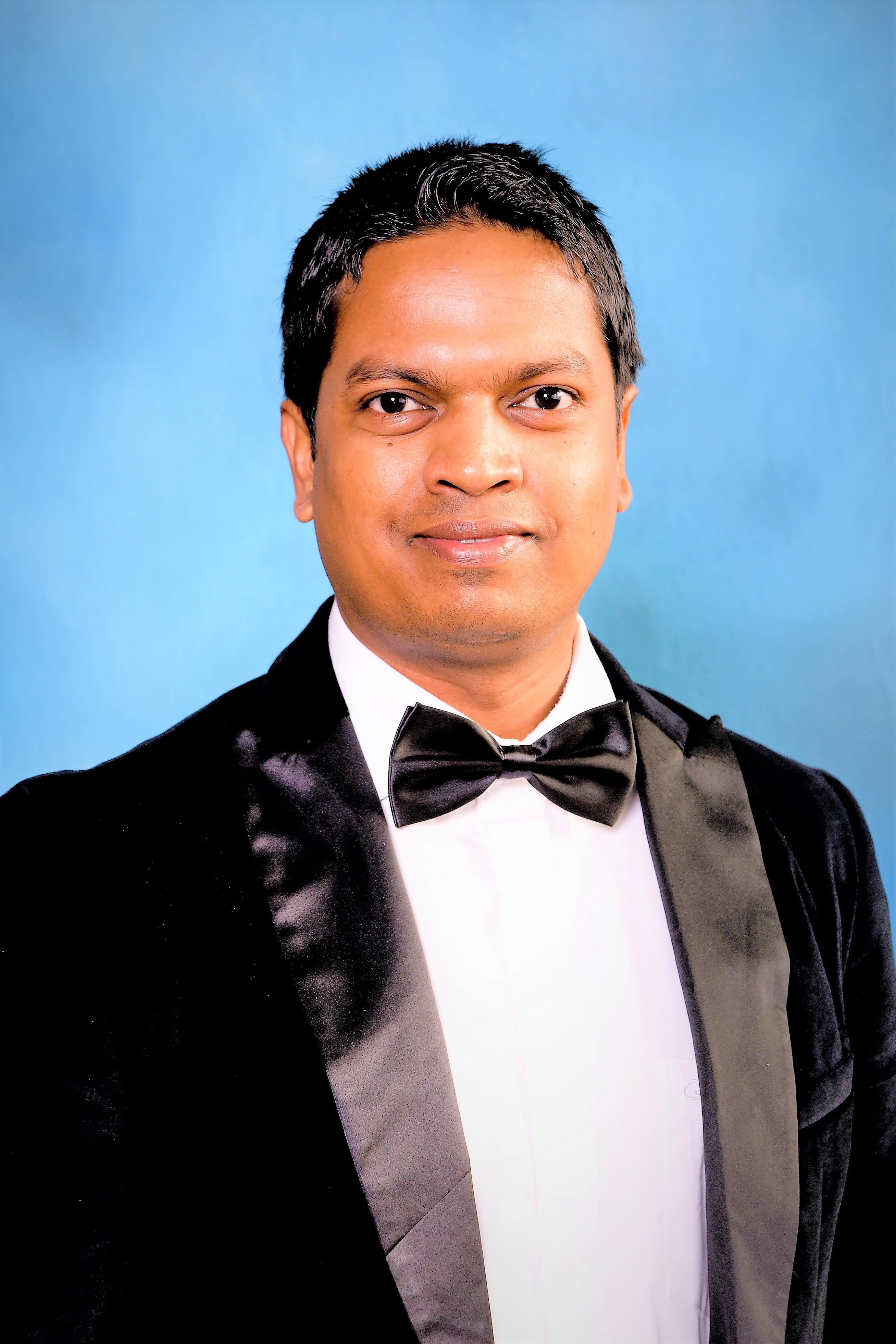}}]{Riadul Islam}
	is currently an assistant professor in the
Department of Computer Science and Electrical Engineering at the University of Maryland, Baltimore County. 
In his Ph.D. dissertation work at UCSC, Riadul
designed the first current-pulsed flip-flop/register that resulted in the 
first-ever one-to-many current-mode clock distribution networks for
high-performance microprocessors. From 2017 to 2019, he was an Assistant
Professor with the University of Michigan, Dearborn MI, USA. 
%From 2007 to 2009, he worked as a full-time faculty member in the Department of Electrical and Electronic
%Engineering of the University of Asia Pacific, Dhaka, Bangladesh. 
He is a senior member of the IEEE, member of the ACM, IEEE Circuits and Systems (CAS) society, the VLSI Systems and Applications Technical 
Committee (VSA-TC) of the IEEE-CAS, and IEEE Solid-State Circuits (SSC) Society. 
%He is a member of the Cybersecurity Center for Research, Education, and Outreach at the UM-Dearborn. 
He holds two US patent and several IEEE/ACM/MDPI/Springer Nature journal and conference publications. 
%in TVLSI, TCAS-I, TCAS-II, TETC, T-ITS, CJECE, JETTA, DAC, ISCAS, MWSCAS, ISQED, and ASICON.  
His  current  research
interests include  digital, analog, and mixed-signal CMOS ICs/SOCs for a 
variety of applications; verification and testing techniques for analog, 
digital and mixed-signal ICs; hardware security; CAN network; CAD tools for design and analysis of
microprocessors and FPGAs; automobile electronics; and biochips. 
%He is an editorial board member of Semiconductor Science and Information Devices journal.
He is an Associate Editor of Springer Circuits, Systems and Signal Processing (CSSP) Journal.
He was a Technical Program Committee (TPC) member of the IEEE/ACM International Conference on Computer-Aided Design (ICCAD 2022), 
%30th ACM Great Lakes Symposium on VLSI (GLSVLSI) 2020
ACM Great Lakes Symposium on VLSI (GLSVLSI 2020, GLSVLSI 2021, GLSVLSI 2022), 57th IEEE/ACM 
Design Automation Conference (DAC) 2020 LBR Session, IEEE Computer Society Annual Symposium on VLSI (ISVLSI) 2021,  and IEEE International Conference on Consumer Electronics (ICCE) 2021.
Riadul is the
recipient of a 2021 NSF ERI award, 2021 Maryland Industrial Partnerships (MIPS) award, and 2021 Maryland Innovation Initiative (MII) award.
\end{IEEEbiography}
%\vspace{10.5cm}

%\input{revision.tex}

\end{document}